\documentclass[10pt,journal,compsoc]{IEEEtran}

\usepackage[T1]{fontenc}		
\usepackage[utf8]{inputenc}		
\usepackage{color}				
\usepackage{graphicx}			
\usepackage{microtype} 			
\usepackage{float}              
\usepackage{caption}

\usepackage{fancyhdr}
\usepackage{authblk}
\usepackage{xcolor,soul,framed} 
\colorlet{shadecolor}{yellow}
\usepackage{subfig,amsmath,cleveref,caption}
\ifCLASSOPTIONcompsoc
  \usepackage[nocompress]{cite}
\else
  \usepackage{cite}
\fi
\ifCLASSINFOpdf
 
\else
 
\fi

\begin{document}

\title{An Audio Synthesis Framework Derived from Industrial Process Control}

\author{Ashwin Pillay}
\affil{University of Mumbai, Mumbai, India}
\affil{\textit {(2016.ashwin.pillay@ves.ac.in)}}

\markboth{Pillay}{An Audio Envelope Generator Derived from Industrial Process Control}

\IEEEtitleabstractindextext{%
\begin{abstract}
Since its conception, digital synthesis has significantly influenced the advancement of music, leading to new genres and production styles. Through existing synthesis techniques, one can recreate naturally occurring sounds as well as generate innovative artificial timbres. However, research in audio technology continues to pursue new methods of synthesizing sounds, keeping the transformation of music constant. This research attempts to formulate the framework of a new synthesis technique by redefining the popular Proportional-Integral-Derivative (PID) algorithm used in feedback-based process control. The framework is then implemented as a Python application to study the available control parameters and their effect on the synthesized output. Further,  applications of this technique as an audio signal and LFO generator, including its potentiality as an alternative to FM and Wavetable synthesis techniques, are studied in detail. The research concludes by highlighting some of the imperfections in the current framework and the possible research directions to be considered to address them.
\end{abstract}}

\maketitle
\IEEEdisplaynontitleabstractindextext
\IEEEpeerreviewmaketitle

\section{Introduction}
\label{S0}
In pursuit of recreating the treasure trove of musical and non-musical sounds found in nature as well as realizing those stemming directly from human creativity, numerous audio synthesis strategies have been devised and developed over time. The additive synthesis technique, originating from the applications of Fourier theory, found use in the earliest synthesizer implementations like the Telharmonium (1897) \cite{weidenaar} and the Hammond Organ (1935) \cite{vail}. On the other hand, subtractive synthesis saw increased popularity in the 1960s and was employed in many popular synthesizers like the Moog Synthesizer\cite{pekonen}.

The Frequency Modulation (FM) Synthesis technique devised by John Chowning \cite{chowning} took audio synthesis into newfound territories, with implementations like Yamaha's DX7 (1983) achieving cult status in the music industry. FM synthesis paved the way for efficiently emulating the complex and dynamic spectra of naturally occurring sounds. Since then, wavetable \cite{bristow} and granular synthesis \cite{roads} techniques have also found wide acceptance in hardware and software synths.

Parallel to the aforementioned developments in audio engineering, the process control industry saw a similar advancement in the proportional-integral-derivative (PID) control strategy. Currently, PID controllers are the most commonly used means of achieving feedback control in industrial plants\cite{lin}. They also find widespread use in automobile control, HVAC systems and robotics.

However, despite the popularity and robustness of PID control in various applications, its usability in the audio domain has not received significant research attention. Consequently, this study documents the application of the PID control strategy in synthesizing musical and non-musical sounds by defining the framework for the PID Synthesis (PIDS) technique. Additionally, an effort is also made to analyze the waveforms and spectrums generated using PIDS, thereby juxtaposing it with existing methods like FM and wavetable synthesis.

\subsection{The PIDS Framework}
\label{S1}

\begin{figure}
\centering
\subfloat[]{\includegraphics[width=0.24\textwidth]{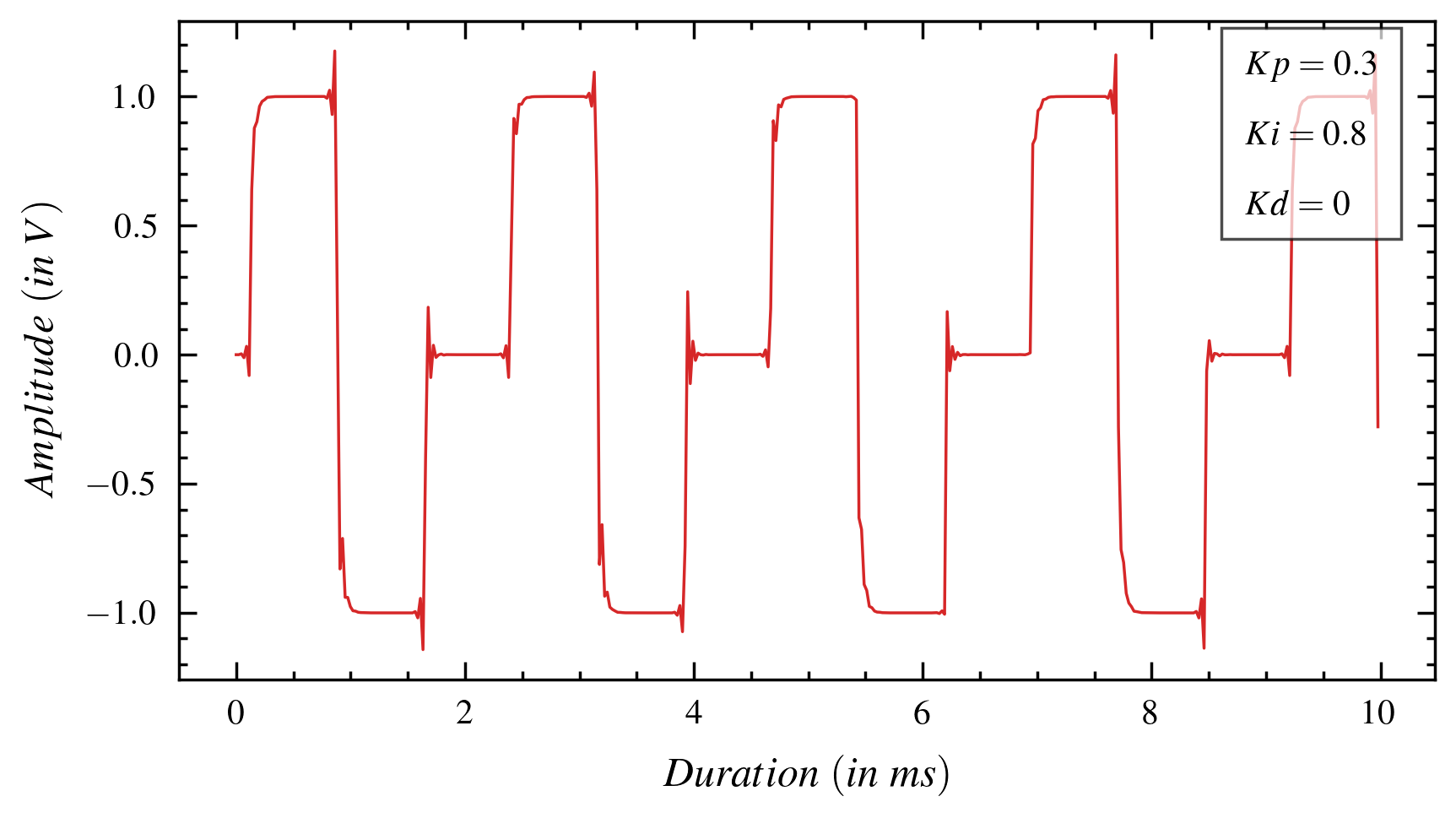}\label{F1a}}
\subfloat[]{\includegraphics[width=0.24\textwidth]{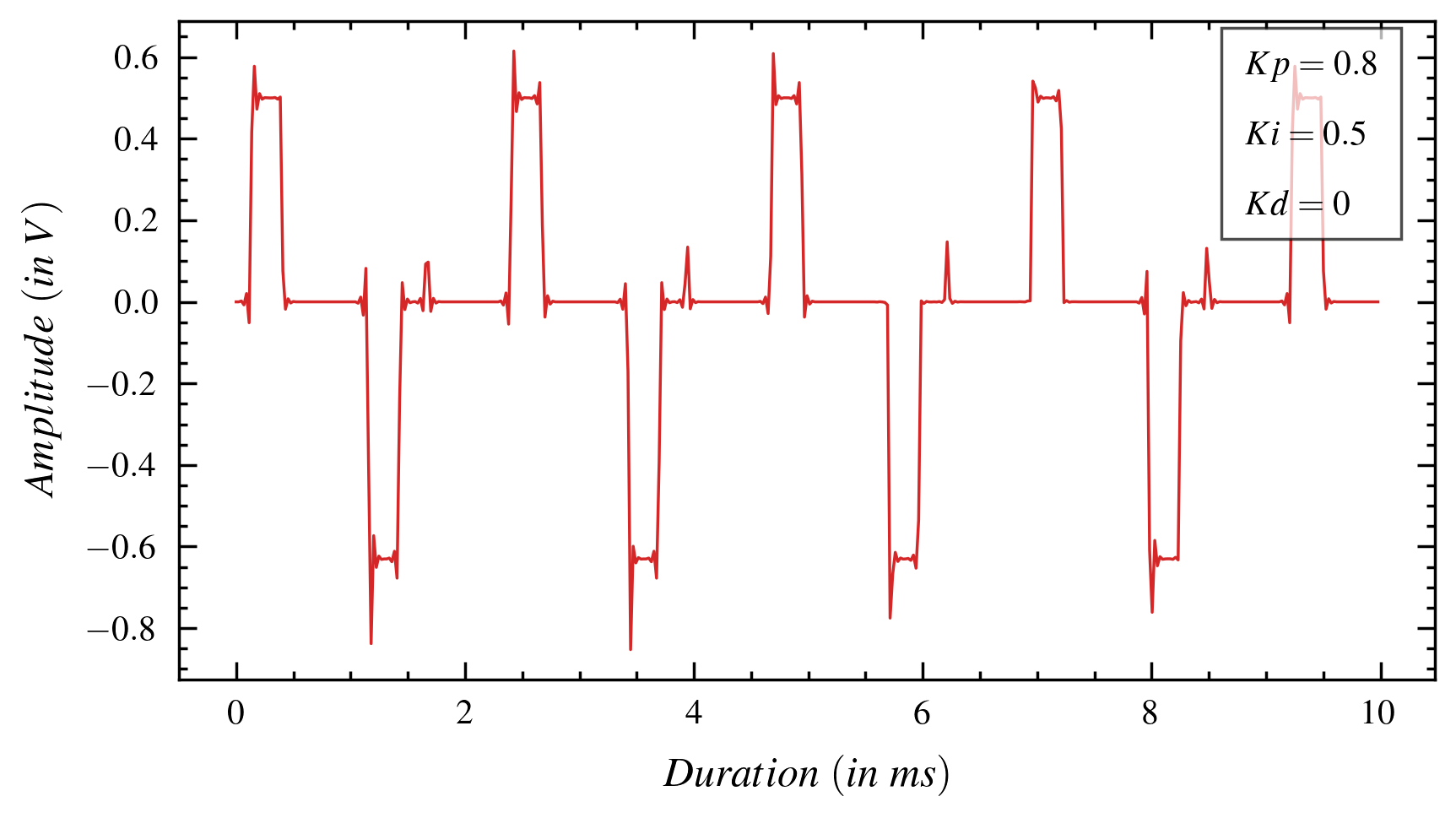}\label{F1b}}\hfill
\subfloat[]{\includegraphics[width=0.24\textwidth]{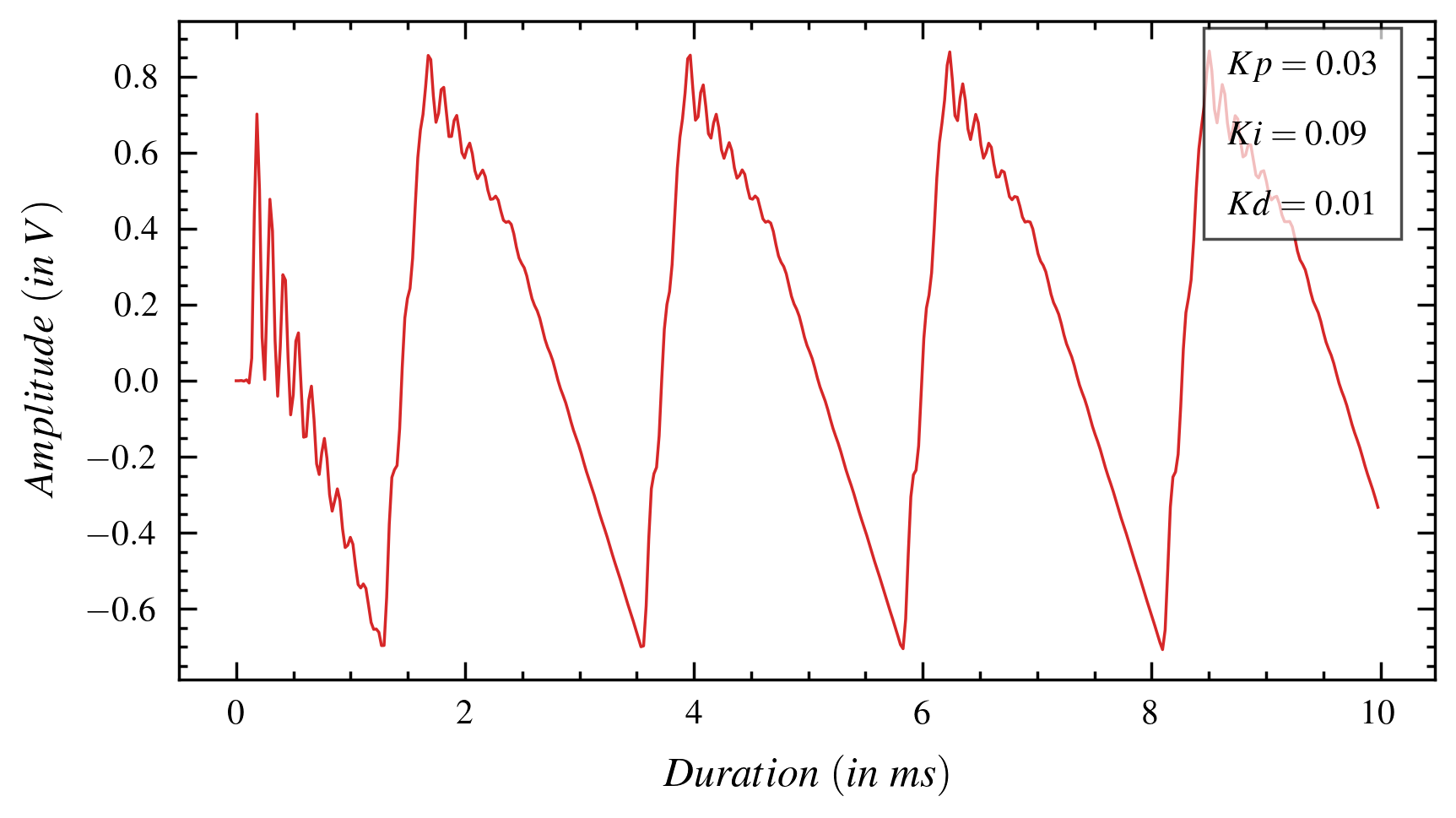}\label{F1c}}
\subfloat[]{\includegraphics[width=0.24\textwidth]{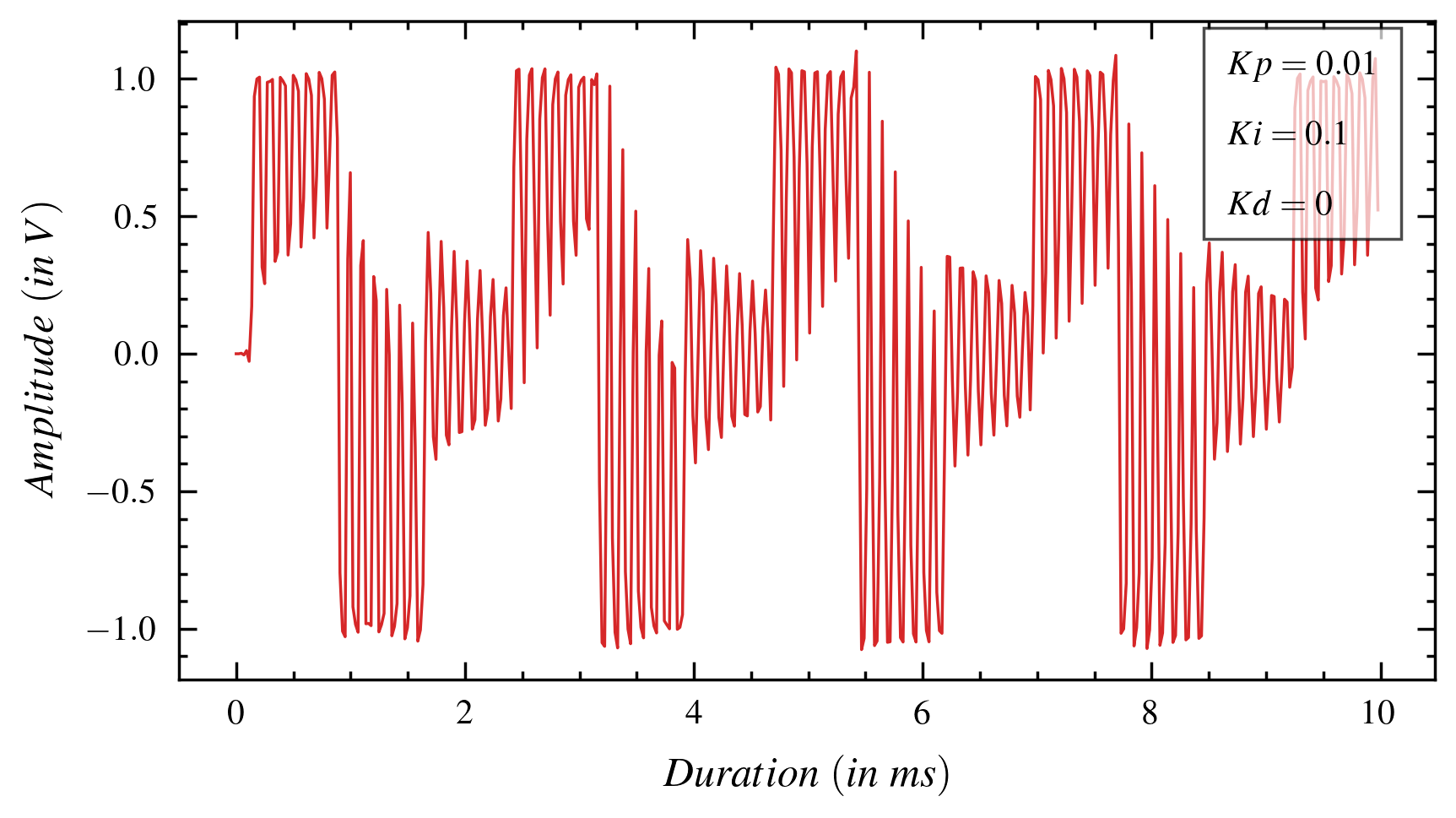}\label{F1d}}\hfill
\caption{Depending on the values of the P, I and D gains set, the PIDS technique can be employed to generate a wide variety of waveforms. If such waveforms are generated at frequencies less than 20 Hz, they can be used as LFOs.}
\label{F1}
\end{figure}

The PID algorithm finds widespread use in various process control systems to perform feedback control, specifically, to bring the value of a process variable (PV) up to a value known as the setpoint (SP).

In most control applications, the SP is either a constant or changes infrequently and is equal to the desired value(s) to which the PV must be set. However, if the SP is allowed to vary as a curve, the PV can be made to follow it around constantly; the nature of its following being determined by the values of $K_{p}$, $K_{i}$ and $K_{d}$; the respective gains of the Proportional (P), Integral (I) and Derivative (D) components.

If the frequency of the SP curve referred hereafter as the "$artist$" is f, the frequency of the resulting PV curve termed as the "$output$" will also be f, granted the values of $K_{p}$ or $K_{i}$ are high enough. In this case, the nature of the $artist$ and the magnitudes of the P, I and D gains (provided as user input or modulated programmatically) control the shape of the $output$. This lays the foundation of PIDS, a technique to generate signals which can be either Low - Frequency Oscillations (LFOs) when f is under 20 Hz or audible waveforms when f is in the audio frequency range. Some of the typical waveforms generated by PIDS are illustrated by \cref{F1}.

However, the PID algorithm does not intrinsically ensure that the $output$ is restricted within the desired threshold. During its process to settle about SP, the PV is susceptible to overshoots. Depending on the PID gains set, such overshoots can assume enormous values. In such cases, the resulting audio signal produced by PIDS is likely to clip (i.e., its level exceeds 0 dB). As a result, the PIDS algorithm restricts the $artist$ range to [-1,1].

Similarly, the range of the $output$ is also contained to [-1, 1]. Any sample above this value is truncated to be within this range. Such truncations can cause undesirable effects in the form of high-frequency harmonics in the signal, leading to issues like aliasing. However, these are addressed in \cref{S3}.

\begin{figure}
\centering
\subfloat[PIDS $output$ with the integral limit set to $\pm5000$]{\includegraphics[width=0.24\textwidth]{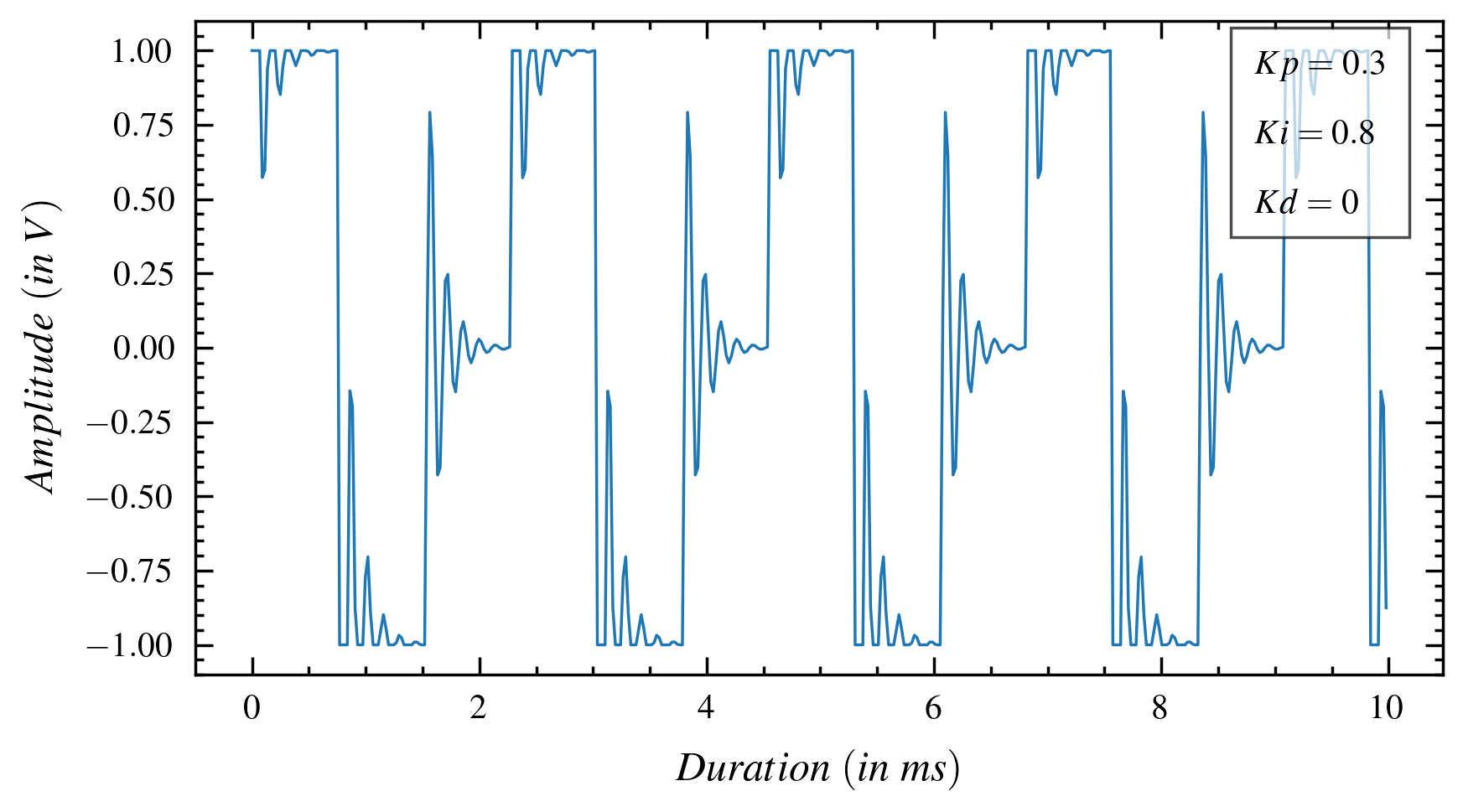}\label{F2a}}
\subfloat[PIDS $output$ with the integral limit set to $\pm1$ (early limiting)]{\includegraphics[width=0.24\textwidth]{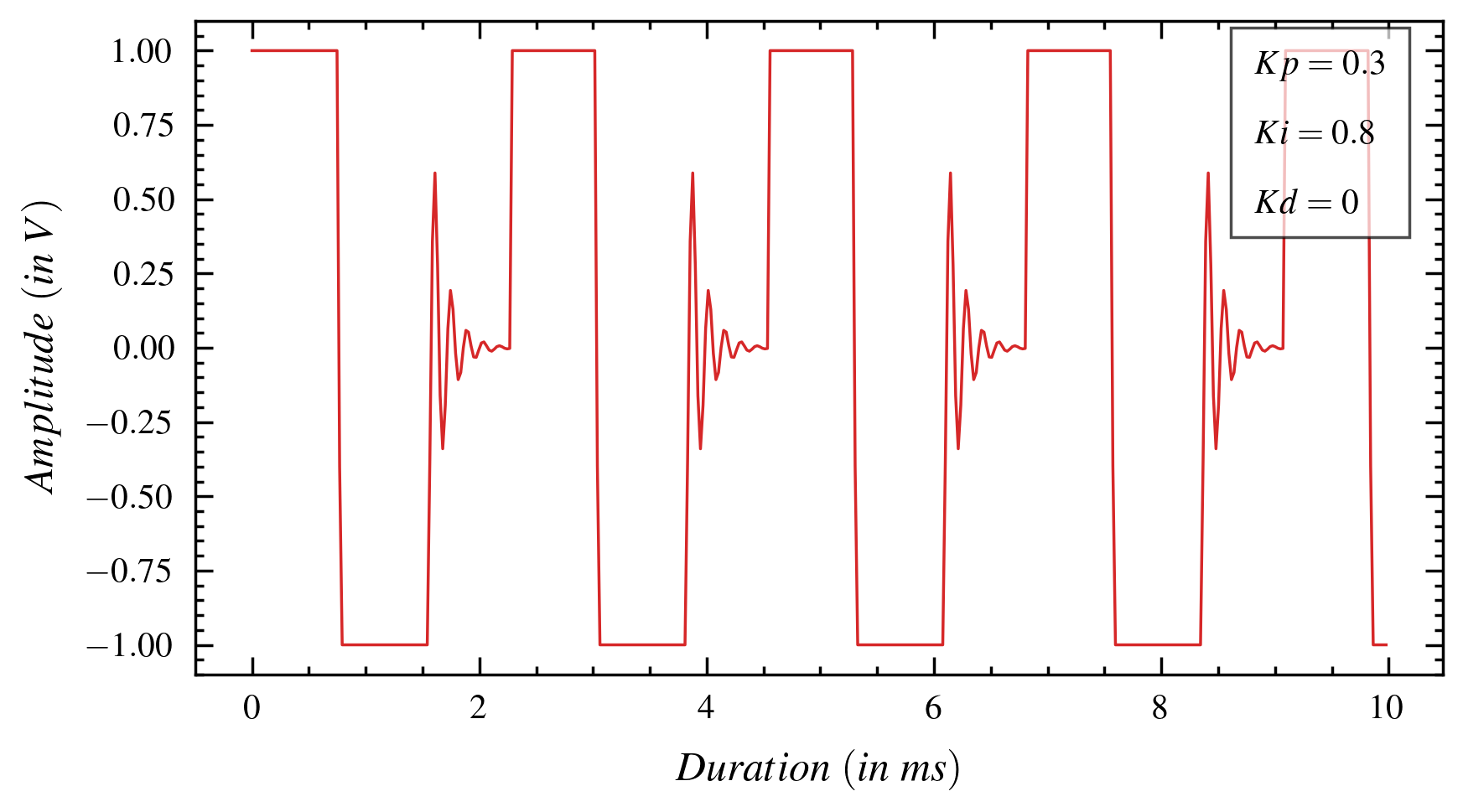}\label{F2b}}\hfill
\caption{As seen in \cref{F2a}, allowing the integral to accumulate to large values lead to oscillations in the $output$ about its highest and lowest levels. This adds high-frequency harmonics to the resulting audio spectrum. In \cref{F2b}, the integral limit is set at a much lower value. The early saturation of the integral clamps the $output$ at the limits, making it more similar to square waves.}
\label{F2}
\end{figure}

Additionally, the error integral may quickly accumulate when the PID algorithm activates the integral mode, and the $K_{i}$ parameter is set to sufficiently high values. Such accumulations can exceed the data type range associated with the integral variable in software synthesizers and can lead to capacitors reaching their charge limits in analogue hardware synthesizers. This saturation condition, known as an integral windup, is generally undesirable in control systems. However, when applied to PIDS, windups can lead to exciting waveforms. Consequently, the PIDS framework accommodates early limiting of the maximum integral value to hasten the windup phenomenon during signal synthesis. The characteristic differences in doing so are depicted in \cref{F2}.

\section{The Artist}
\label{S2}

In PIDS, the $artist$ determines the instantaneous SP against which the PID algorithm generates an $output$ sample. At higher values of $K_{p}$ and $K_{i}$, the frequency of this SP curve directly sets the frequency of the generated waveform. To synthesize an audio signal of frequency f, the $artist$ wavelength required in terms of sample count is given by \cref{E1}.

\begin{equation}
wavelength = \frac{sampling\_rate}{f}
\label{E1}
\end{equation}

The PID scheme places no restrictions on the type of $artists$ that can be supplied to it. However, to ensure that meaningful audio signals are generated for all possible inputs provided by a user, a PIDS implementation can define a set of skeletons that the $artist$ can assume. Here, the degree of freedom offered to the user is in terms of selecting one of the available skeletons and then varying its subtleties to control the audio waveform produced as desired.

To provide user control for varying the subtleties, the concept of breakpoints is introduced. Breakpoints are the set of x- and y-coordinates through which the $artist$ must strictly pass through during a single wave cycle. For each breakpoint, the x-coordinate can assume any value in [0, 1] and represents the relative position of that point in the wave cycle. On the other hand, y-coordinates can have values within [-1, 1] and describe the amplitude of the $artist$ at that point. The first $(0, y_{0})$ and last breakpoint $(0, y_{n})$ must be the same to ensure continuity of the $artist$ across multiple wave cycles. However, it may not be validated in practice if the user prefers such sharp changes to contribute towards the synthesized $output$. The manufacturer must place suitable anti-aliasing techniques to produce clean audio signals in these situations, as discussed in \cref{S3}. The remaining breakpoints are positioned through user input. By placing these intermediate breakpoints suitably, the finer aspects of the $artist$ in terms of its amplitude, duty cycle and baseline may be controlled.

Additionally, how the $artist$ traverses the intervals between consecutive breakpoints determines its skeleton (or shape). Conventionally, the PIDS framework abstracts the skeleton generation process from the user. Instead, they are provided with the option of selecting one of the available skeletons, and the algorithm will ensure that the resulting $artist$ produced will pass through all the set breakpoints. Except for some exceptional cases, the $artists$ thus created will have discontinuities at the intermediate breakpoints. In theory, there can be many such breakpoints: the higher their count, the better the flexibility in controlling the $artist$ shape. However, in practice, the sampling rate of synthesis constrains the highest $output$ frequency that can be produced while ensuring the $artist$ passes through all the set breakpoints; this upper limit may be obtained from \cref{E2}.

\begin{equation}
f_{max} = \frac{sampling\_rate}{\#\ of\ breakpoints}
\label{E2}
\end{equation}

\subsection{Implementational Choices}
\label{S21}

In the following section, some of the common $artists$ compatible with PIDS will be discussed.

\subsubsection{Linear}
\label{S211}

Linear $artists$ (made up of linear skeletons) are the stock option provided by the PIDS framework. It is obtained by performing linear interpolation between consecutive breakpoints. Apart from the relative ease of generating linear $artists$, the per-sample change in the SPs produced is gradual. This ensures that the $output$ shape does not deviate from the $artist$ to a great extent. Additionally, it also reduces the possibility of sharp changes being present in the $output$. Therefore, the resulting audio signals produced contain much lesser high-frequency components than for other $artist$ types at the same audio frequency.

Consequently, implementing anti-aliasing for linear $artists$ is relatively simple in most cases. However, linear $artists$ may undergo a significant change in gradient at the breakpoints. While the PID algorithm negates the effect of these discontinuities to a great extent, at higher values of $K_{p}$ and $K_{i}$, it may be required that the anti-aliasing technique employed addresses their undesired effects.

In terms of user control, various $artist$ shapes can be derived from the linear skeleton by placing breakpoints accordingly. Notable ones include the triangle and sawtooth waveforms. Additionally, their duty cycle and baseline can also be varied.

\begin{figure}[b]
\centering
\subfloat[]{\includegraphics[width=0.24\textwidth]{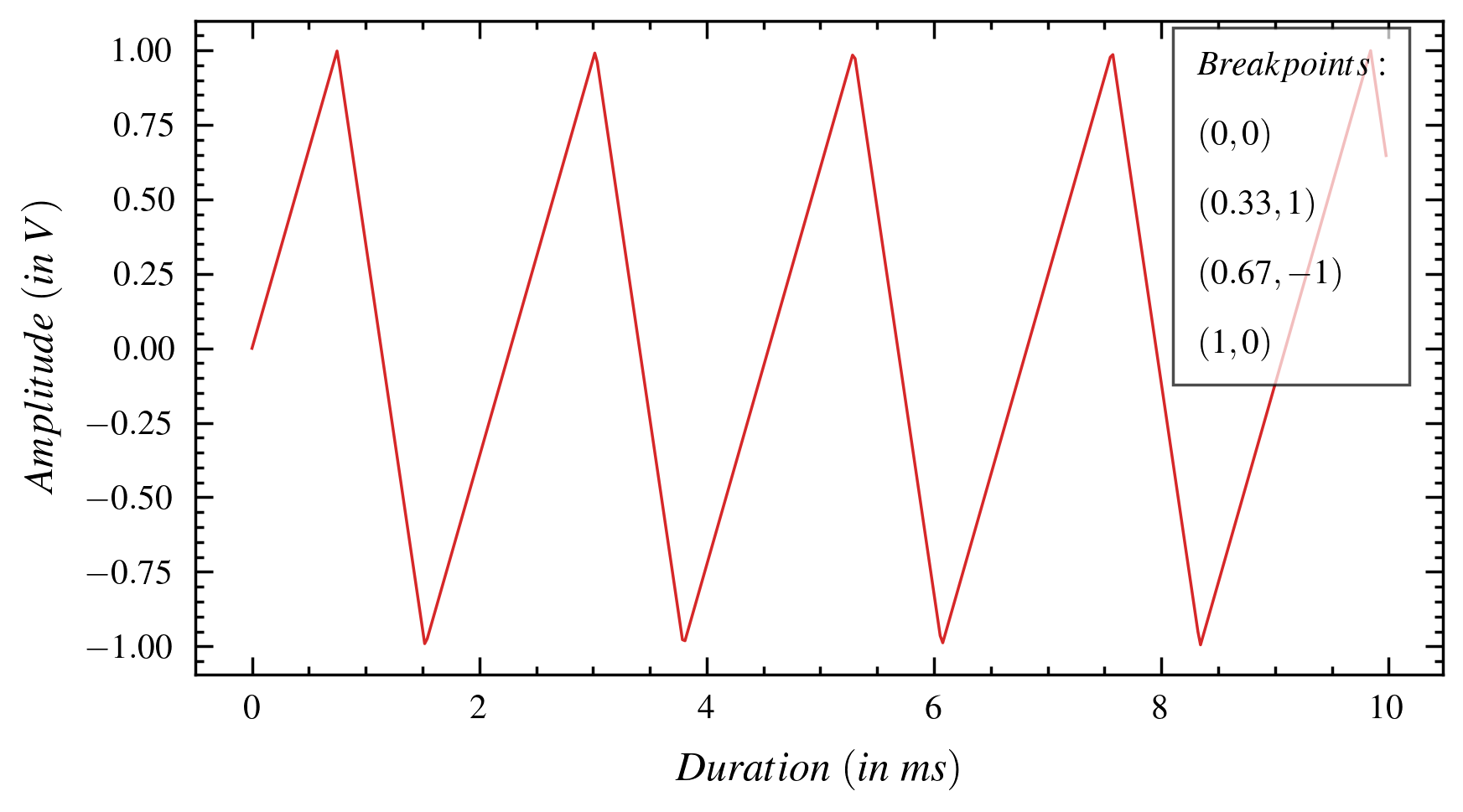}\label{F3a}}
\subfloat[]{\includegraphics[width=0.24\textwidth]{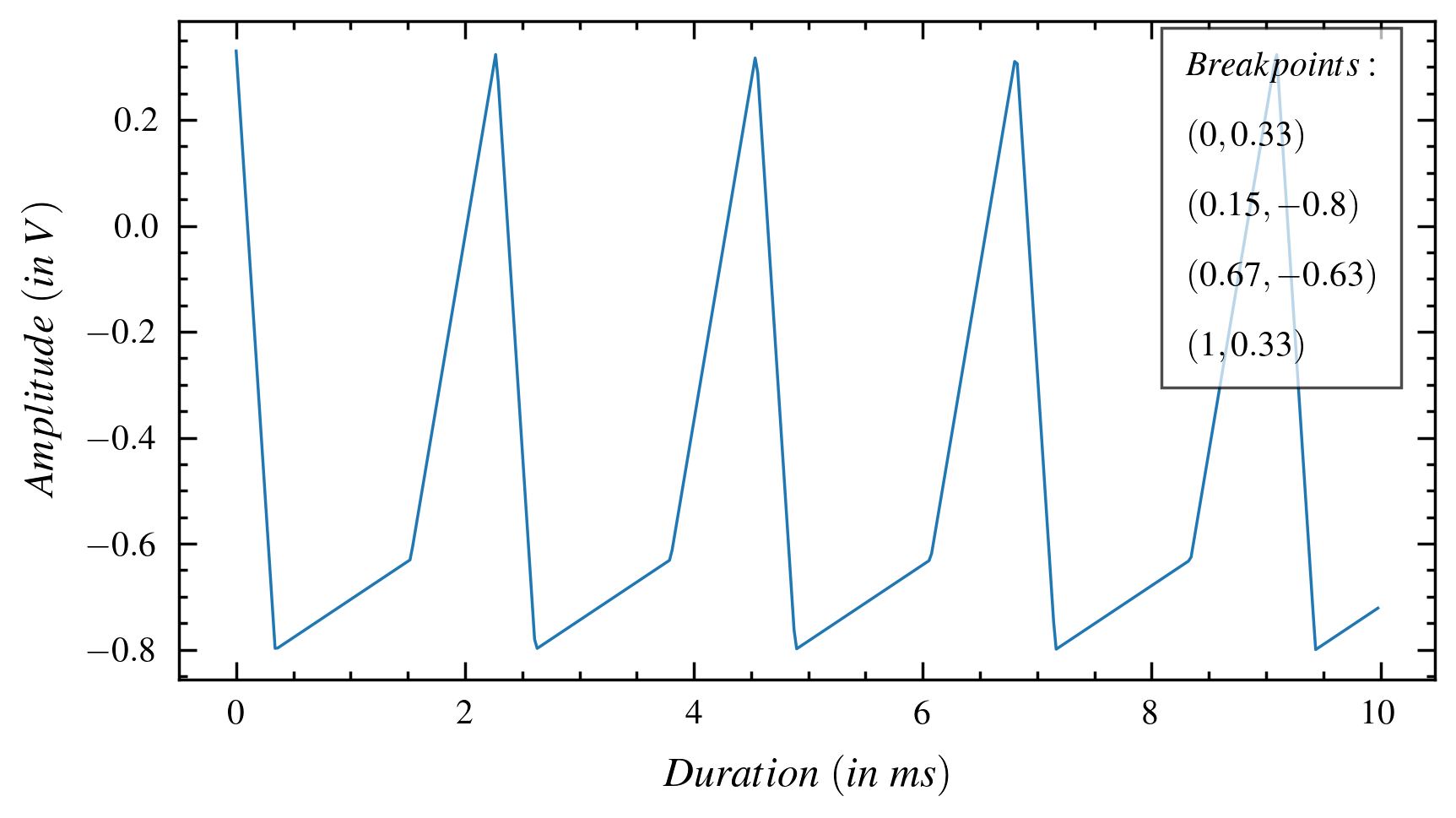}\label{F3b}}\hfill
\subfloat[]{\includegraphics[width=0.24\textwidth]{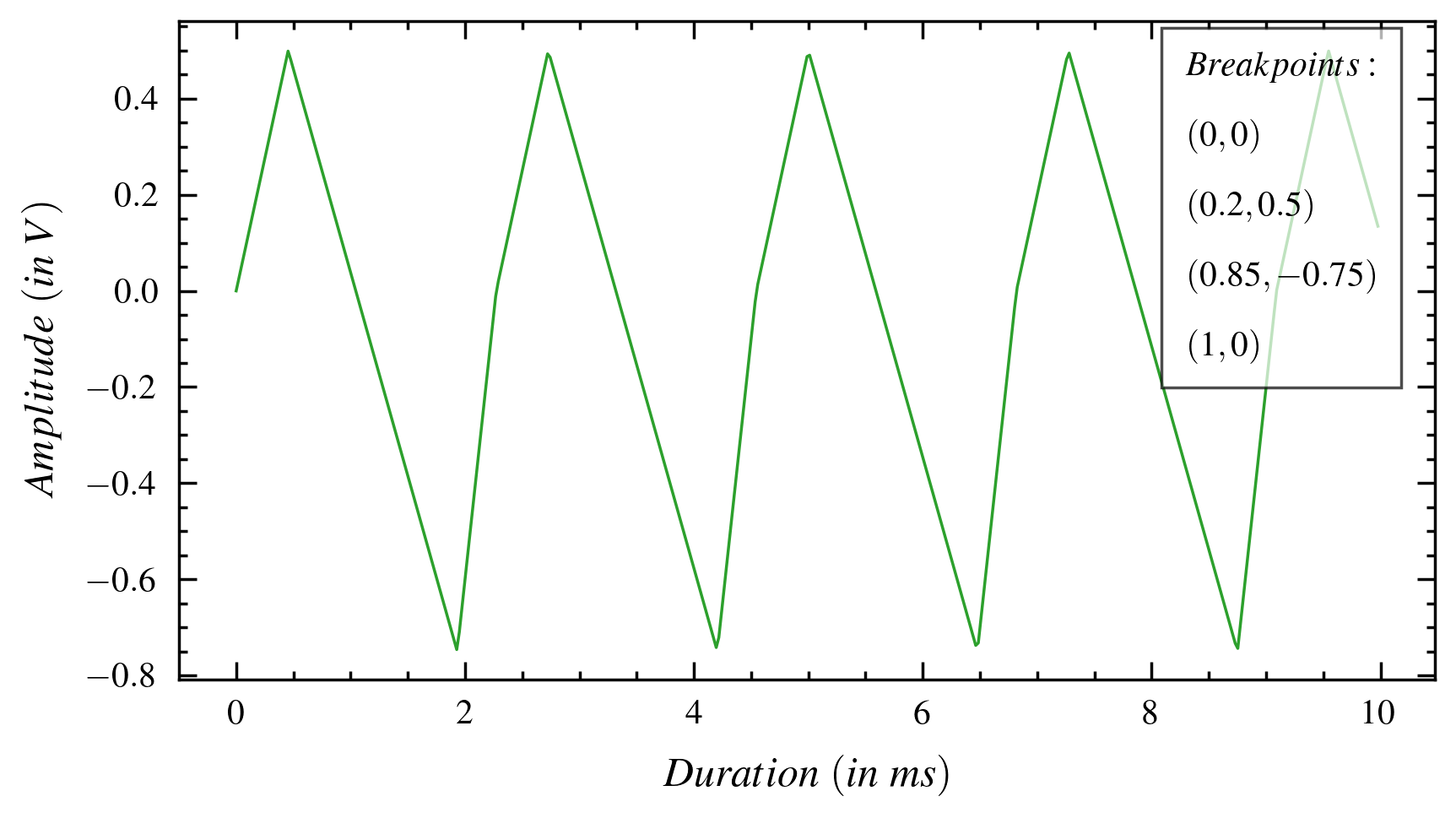}\label{F3c}}
\subfloat[]{\includegraphics[width=0.24\textwidth]{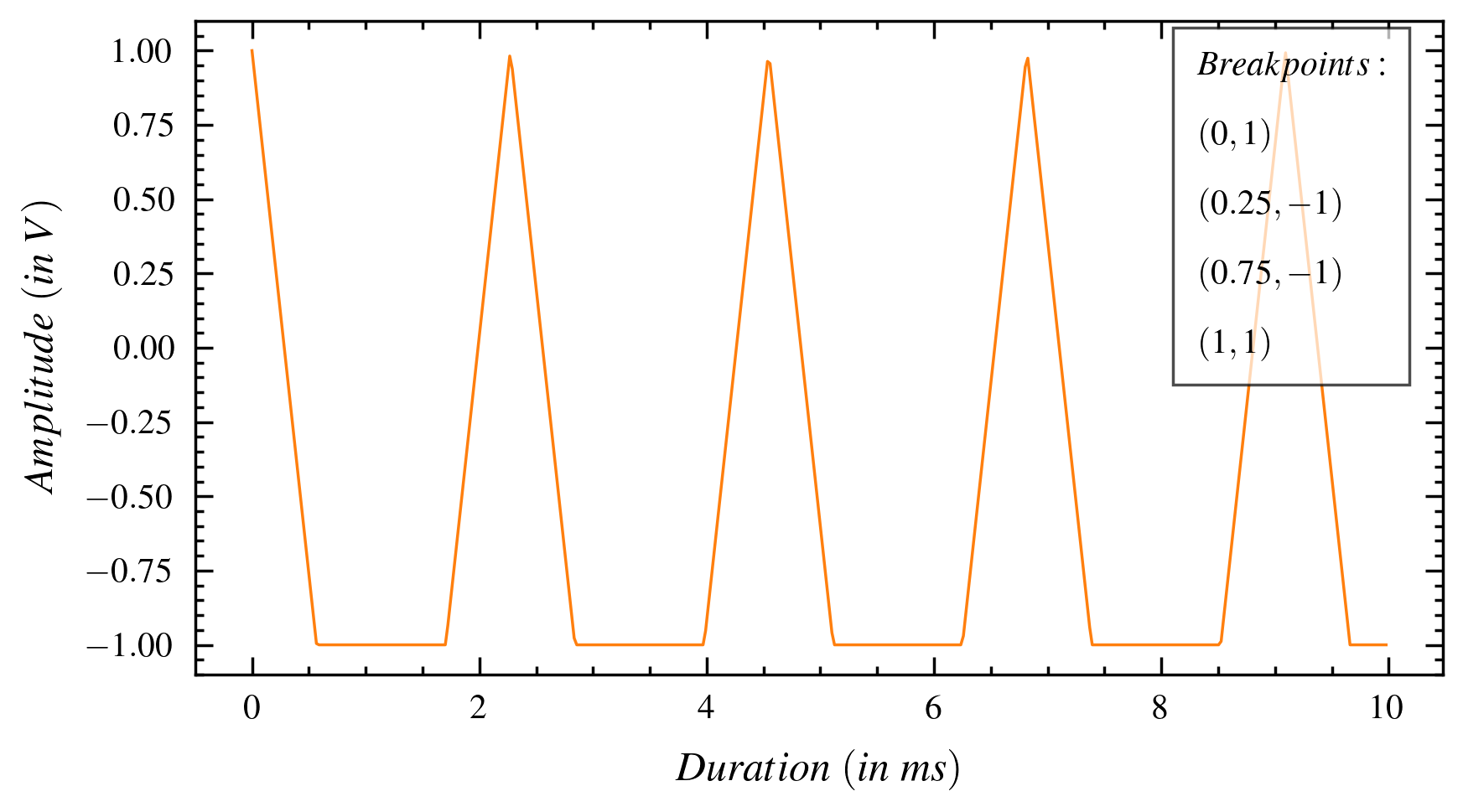}\label{F3d}}\hfill
\caption{Linear $artists$ are obtained through linear interpolation between consecutive breakpoints. They are simple to generate and are versatile, leading to shapes like triangle and sawtooth waves or their different combinations.}
\label{F3}
\end{figure}

\subsubsection{Step}
\label{S212}

Step skeletons are another convenient alternative to generate $artists$ while providing considerable control to the user. Unlike linear $artists$, the SP remains constant between breakpoints in step $artists$. Hence, the PID algorithm has more liberty in setting the value of $output$ samples when compared to linear $artists$. However, the setpoints can change sharply at the breakpoints, leading to discontinuities in the step $artist$. For higher values of $K_{p}$ and $K_{i}$, the $output$ thus produced may also consist of these discontinuities. Hence, appropriate anti-aliasing must account for the infinite high frequencies added by them in the audio signal.

Step $artists$ result in square waveforms. Therefore, by setting the position of breakpoints, the user essentially controls the $artist$ duty cycle, which influences the $output$'s overall shape.

\begin{figure}
\centering
\subfloat[]{\includegraphics[width=0.24\textwidth]{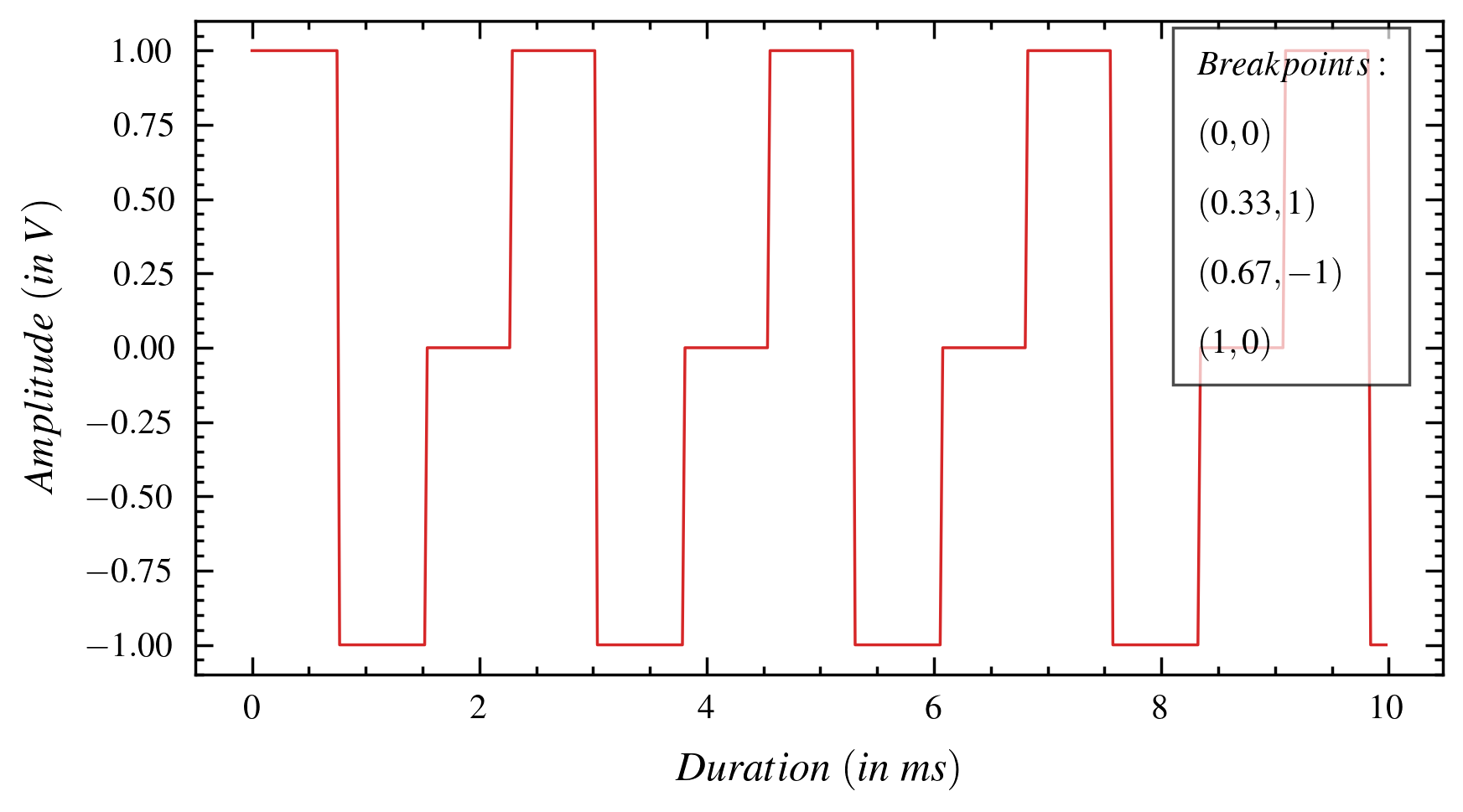}\label{F4a}}
\subfloat[]{\includegraphics[width=0.24\textwidth]{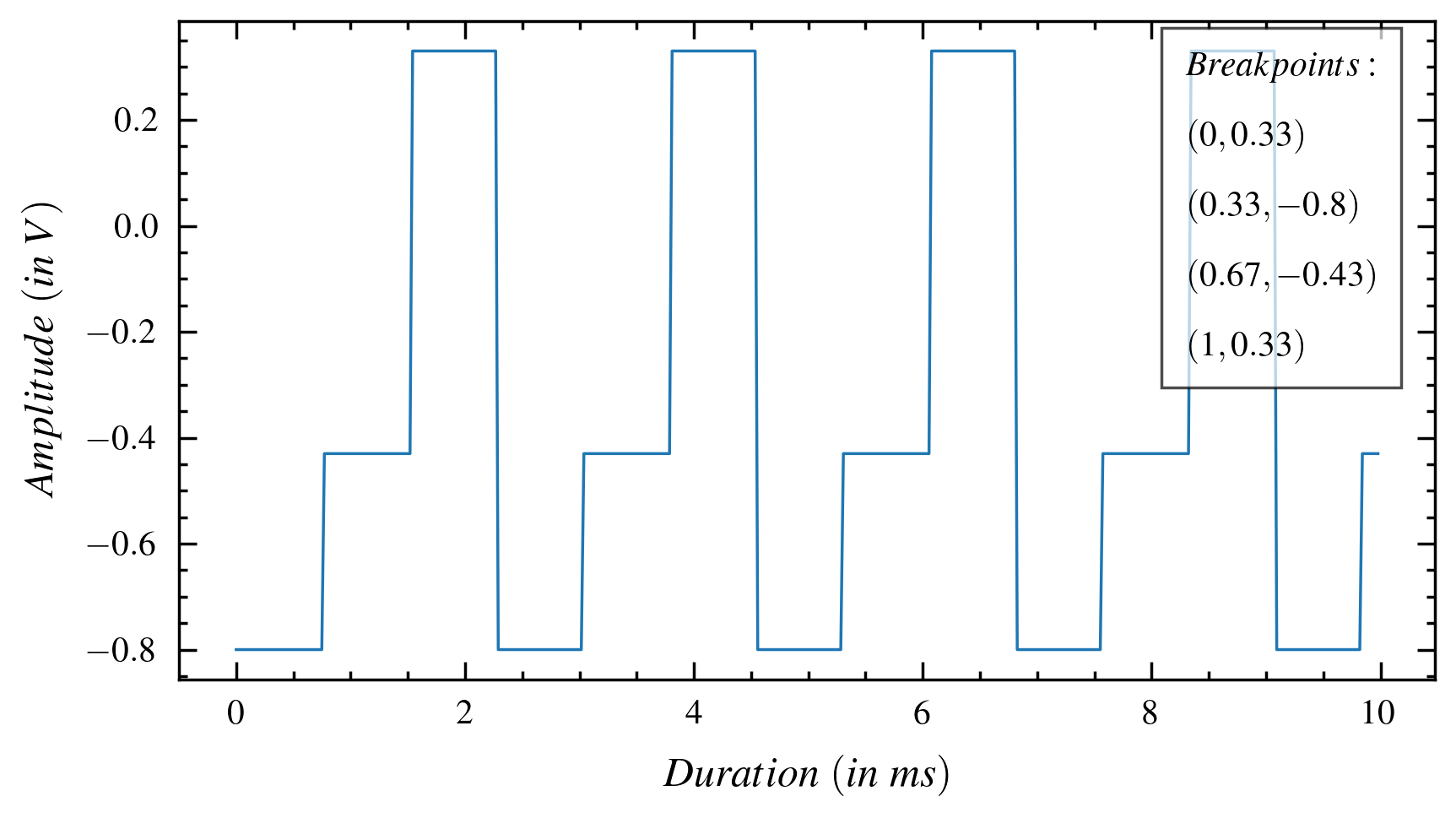}\label{F4b}}\hfill
\subfloat[]{\includegraphics[width=0.24\textwidth]{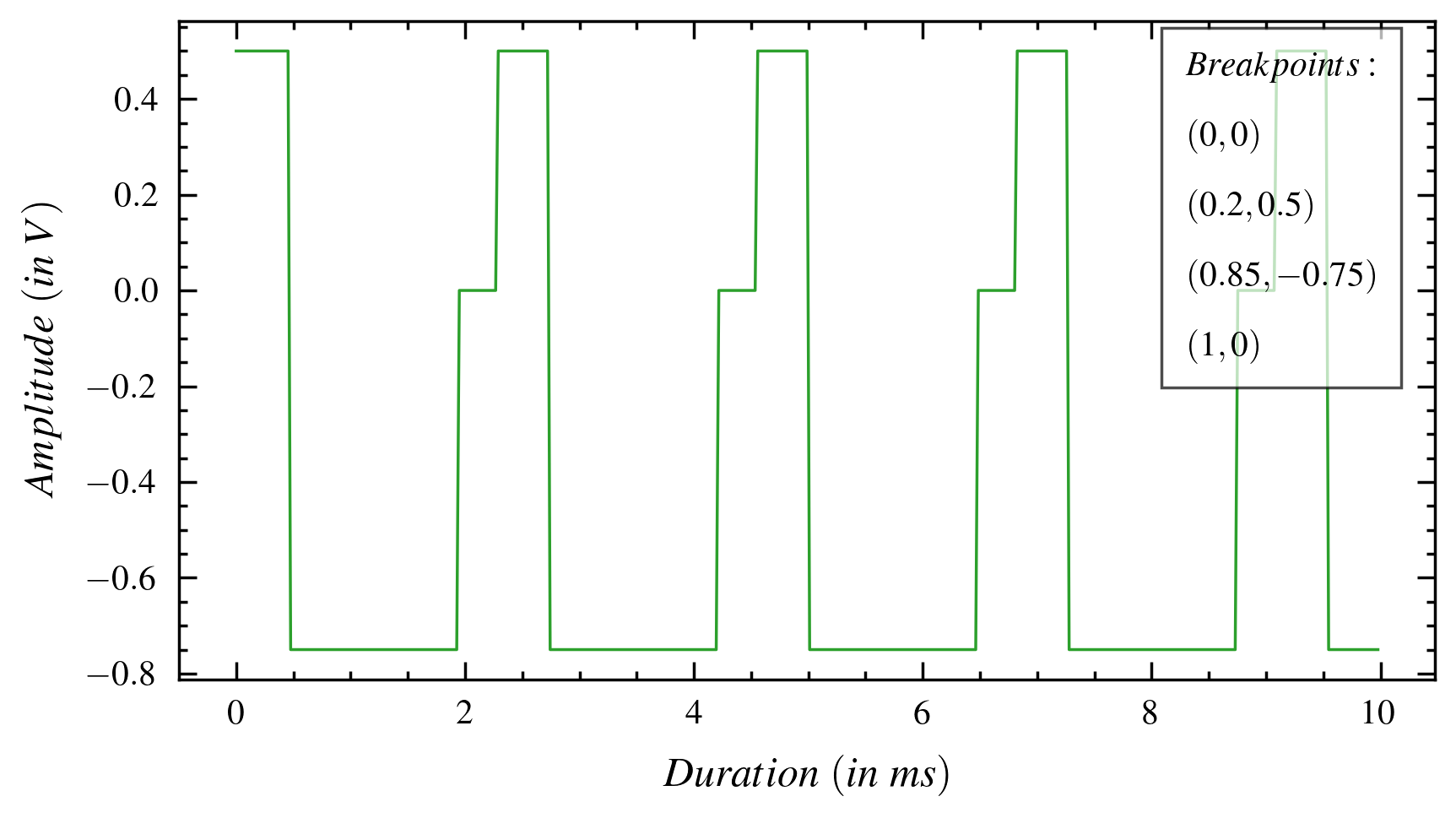}\label{F4c}}
\subfloat[]{\includegraphics[width=0.24\textwidth]{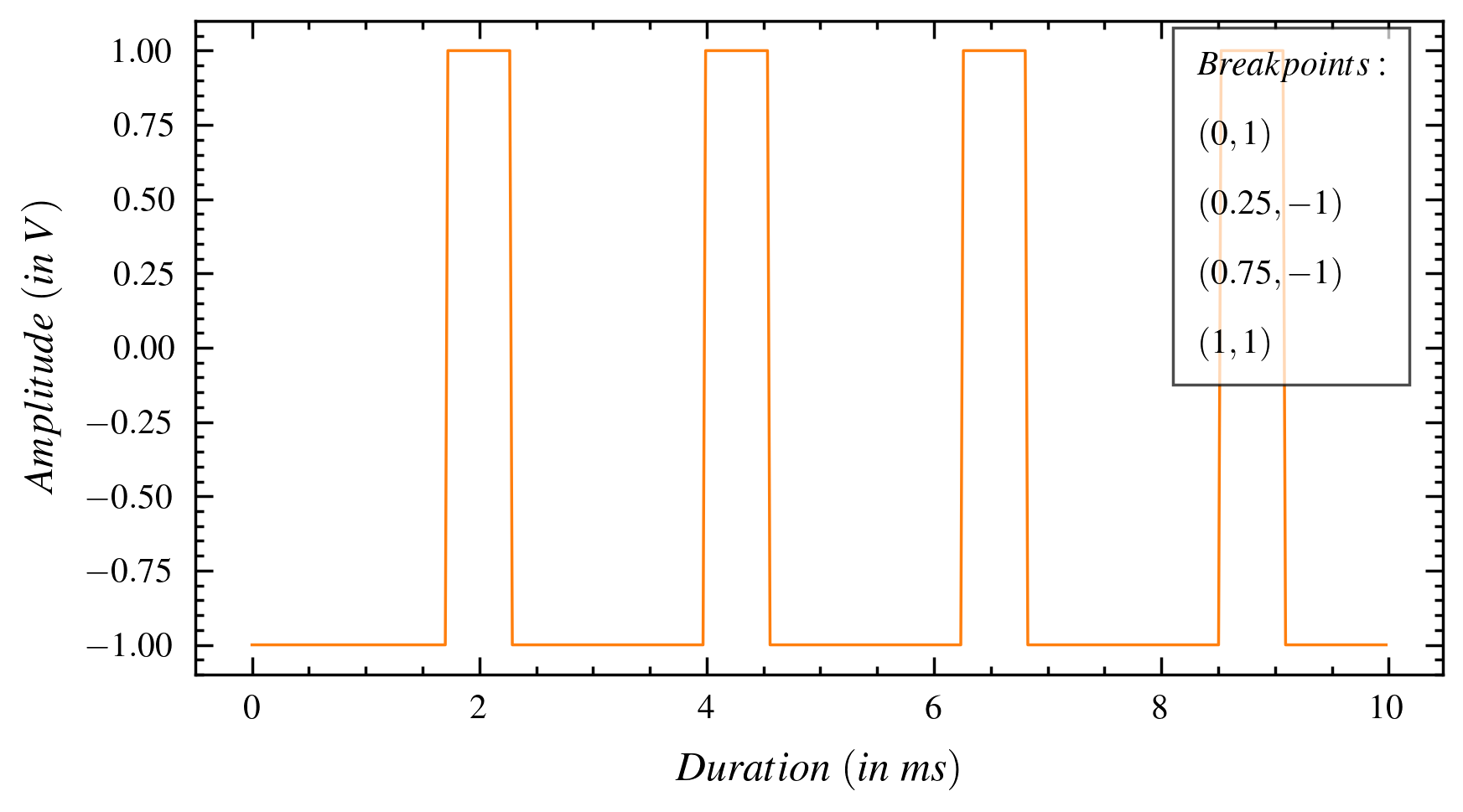}\label{F4d}}\hfill
\caption{Compared with linear $artists$, step $artists$ provide more control to the PID controller in shaping the $output$ waveform. They resemble square waves of varying baselines, amplitudes or duty cycles with a potentially sharp change in values at the breakpoints and constant otherwise.}
\label{F4}
\end{figure}

\subsubsection{Sinusoidal}
\label{S213}

\begin{figure}[b]
\centering
\includegraphics[width=0.48\textwidth]{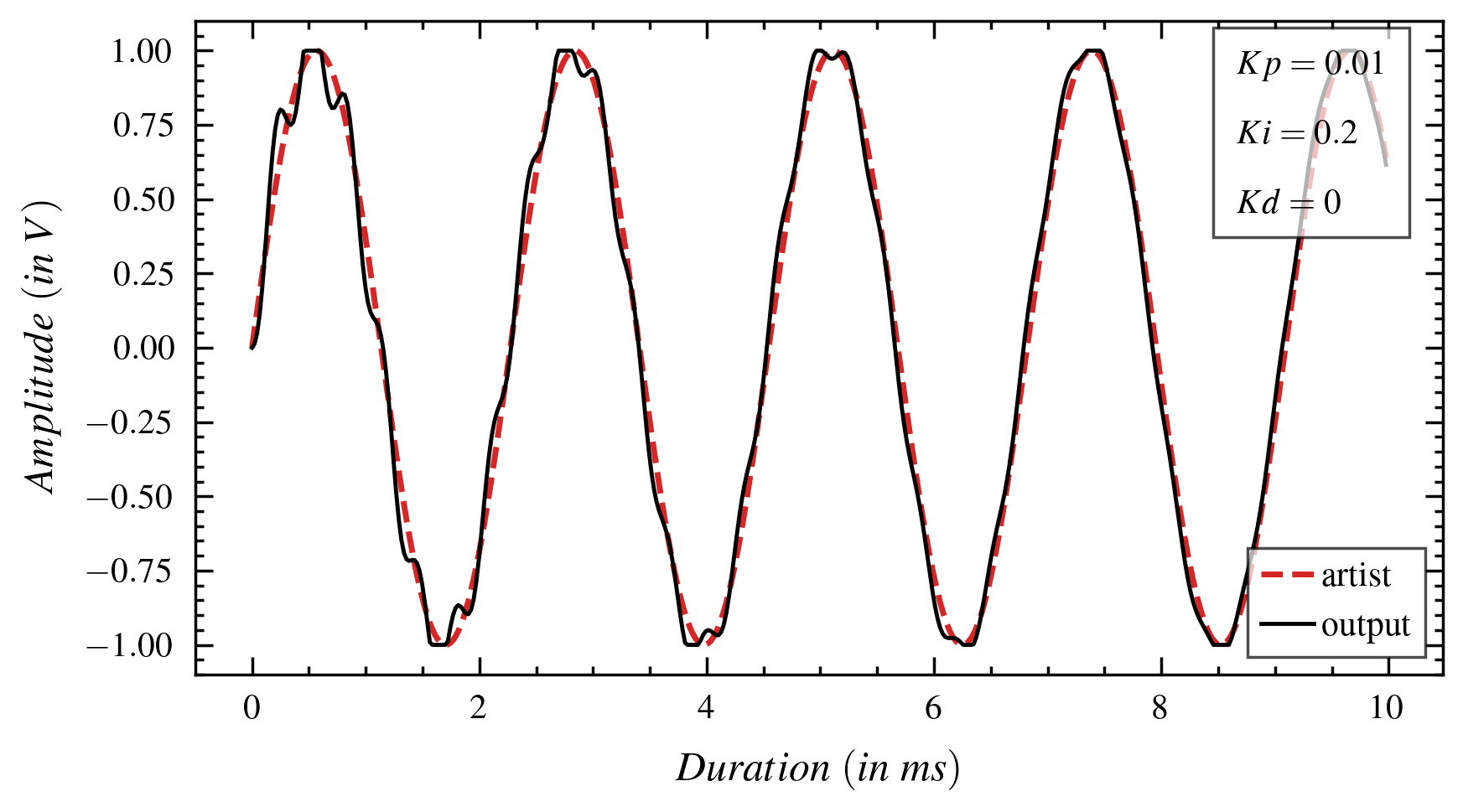}
\caption{Unlike linear and step, sine $artists$ are continuous across breakpoints. As a result, their configurability is restricted to controlling the amplitude and phase alone. However, having a single frequency, sine $artists$ lead to PIDS $output$ that can be easily anti-aliased.}
\label{F5}
\end{figure}

Unlike linear and step $artists$, sinusoidal $artists$ are continuous. Since sine waves contain no harmonics, the $output$ synthesized using sinusoidal $artists$ is less prone to aliasing at higher audio frequencies than its discontinuous counterparts. However, the concept of setting breakpoints is ineffective in the case of sine $artists$. Therefore, the tradeoff for simpler anti-aliasing is a marked decrease in configurability. While facilities can be provided to control the amplitude and phase (useful when using more than 2 PIDS together) of the sine waveform, the inability to set breakpoints might make sinusoidal $artists$ less desirable when compared to some of the discontinuous alternatives.

Additionally, there exist other $artist$ types that can be used with PIDS. For example, these may consist of skeletons of the inverse-exponential and parabolic types. However, they will not be discussed in detail as part of this research.

\section{Anti-Aliasing}
\label{S3}

The $output$ synthesized by PIDS against most of the aforementioned $artist$ types contains high-frequency harmonic and inharmonic components. While these components may lead to natural-sounding or "organic" timbres, some of their frequencies can exceed the Nyquist frequency set by the sampling rate. Such components get folded over and become aliased by lower frequency components on the other side of the axis of symmetry (given by the Nyquist frequency).

Subjectively, aliasing may be a desirable effect in terms of enriching the spectrum of sounds produced. However, the properties of audio signals containing aliased components depend on the sampling rate to a great extent. Hence the same signal might sound different when played by different audio players. Therefore, the PIDS framework incorporates anti-aliasing to keep synthesis solely dependent on the PID algorithm and consistent across audio playbacks.

\begin{figure}[b]
\centering
\subfloat[FFT performed on a step $artist$]{\includegraphics[width=0.24\textwidth]{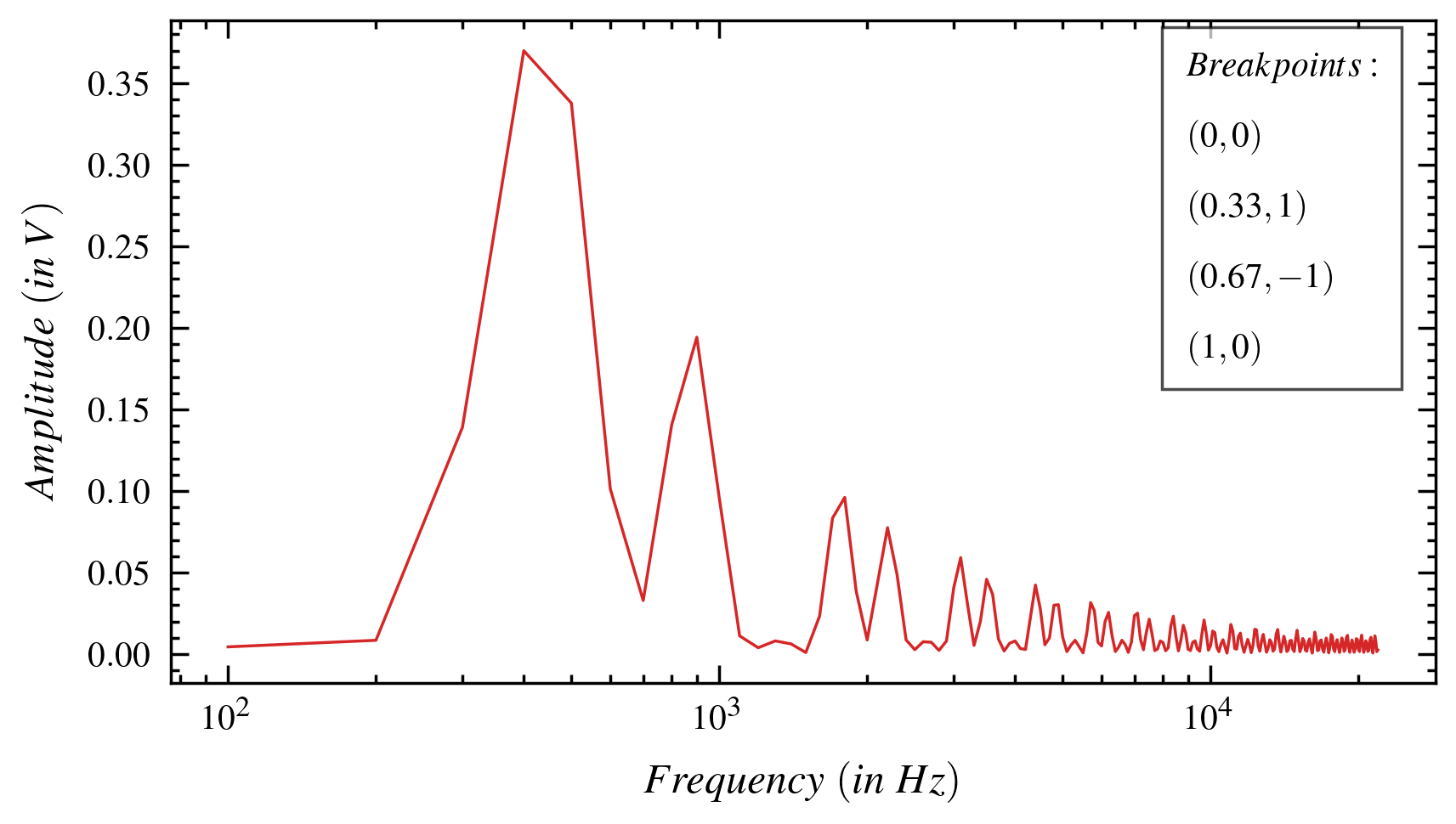}\label{F6a}}
\subfloat[FFT performed on a linear $artist$)]{\includegraphics[width=0.24\textwidth]{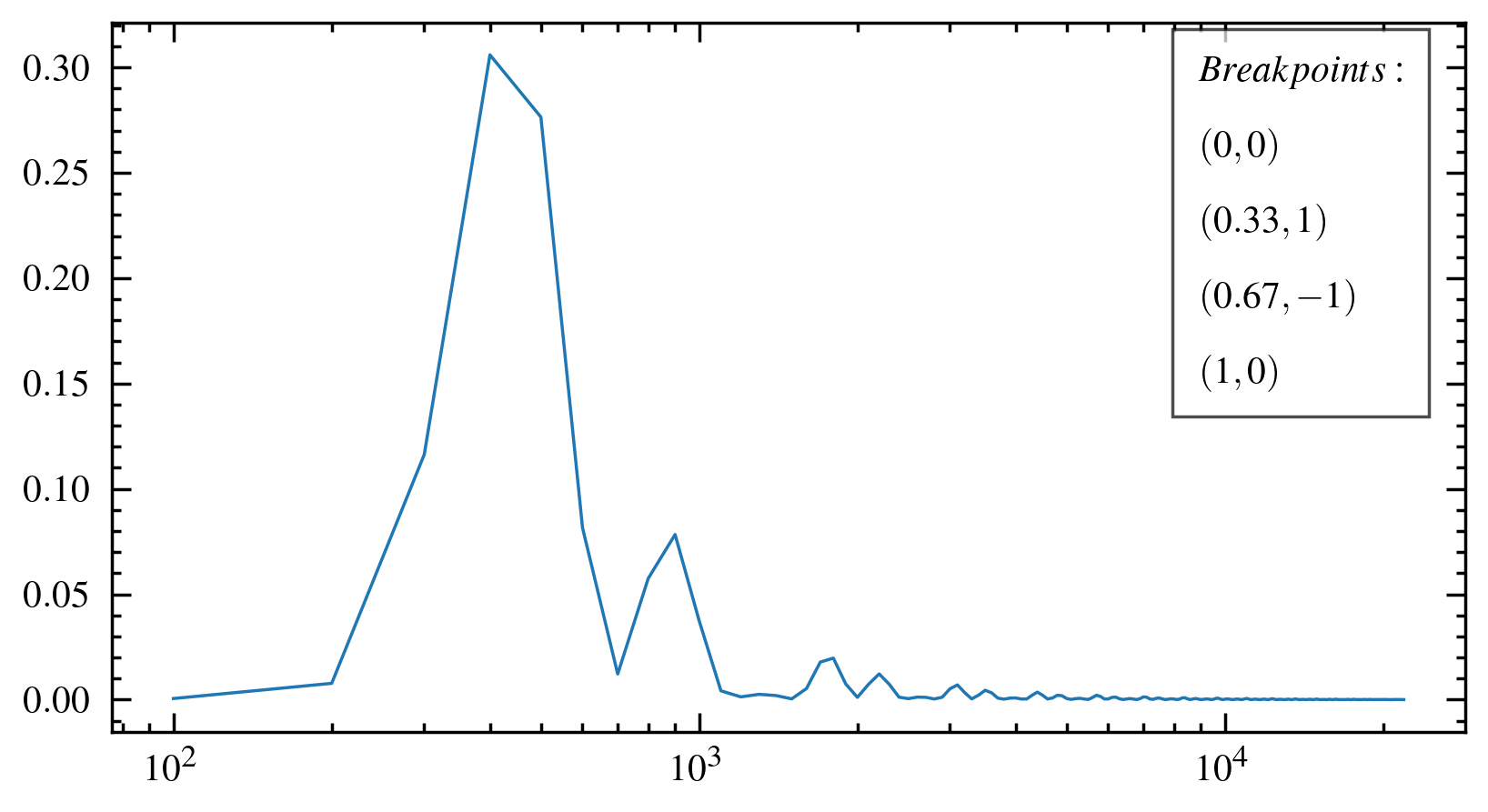}\label{F6b}}\hfill
\caption{Discontinuous $artists$ introduce high-frequency harmonics through sudden changes in amplitude or slope. When the $output$ is synthesized using these $artists$, the high-frequency components of the resultant audio signals get folded over about the Nyquist frequency and lead to the undesirable problems of aliasing.}
\label{F6}
\end{figure}

Especially in the case of discontinuous $artists$, features like a step change in amplitude (occurring in step $artists$) and sudden change in slope (occurring in linear $artists$) induce infinite harmonics that slowly decay to inaudible amplitudes, as seen in \cref{F6}. Therefore, the problem of aliasing appears right from the $artist$ generation process itself. The phase-lag introduced by the PID algorithm at lower values of $K_{p}$ or when the I mode is activated moderates the effects of these aliased $artist$ components to a great extent. However, PID induces high-frequency components of its own while generating $output$ samples against the $artist$. Some of the common causes may be oscillations produced about the SP at higher values of $K_{i}$ and the $output$ getting truncated in the [-1, 1] range at higher values of $K_{p}$.

\begin{figure}
\centering
\subfloat[Aliased Signal]{\includegraphics[width=0.24\textwidth]{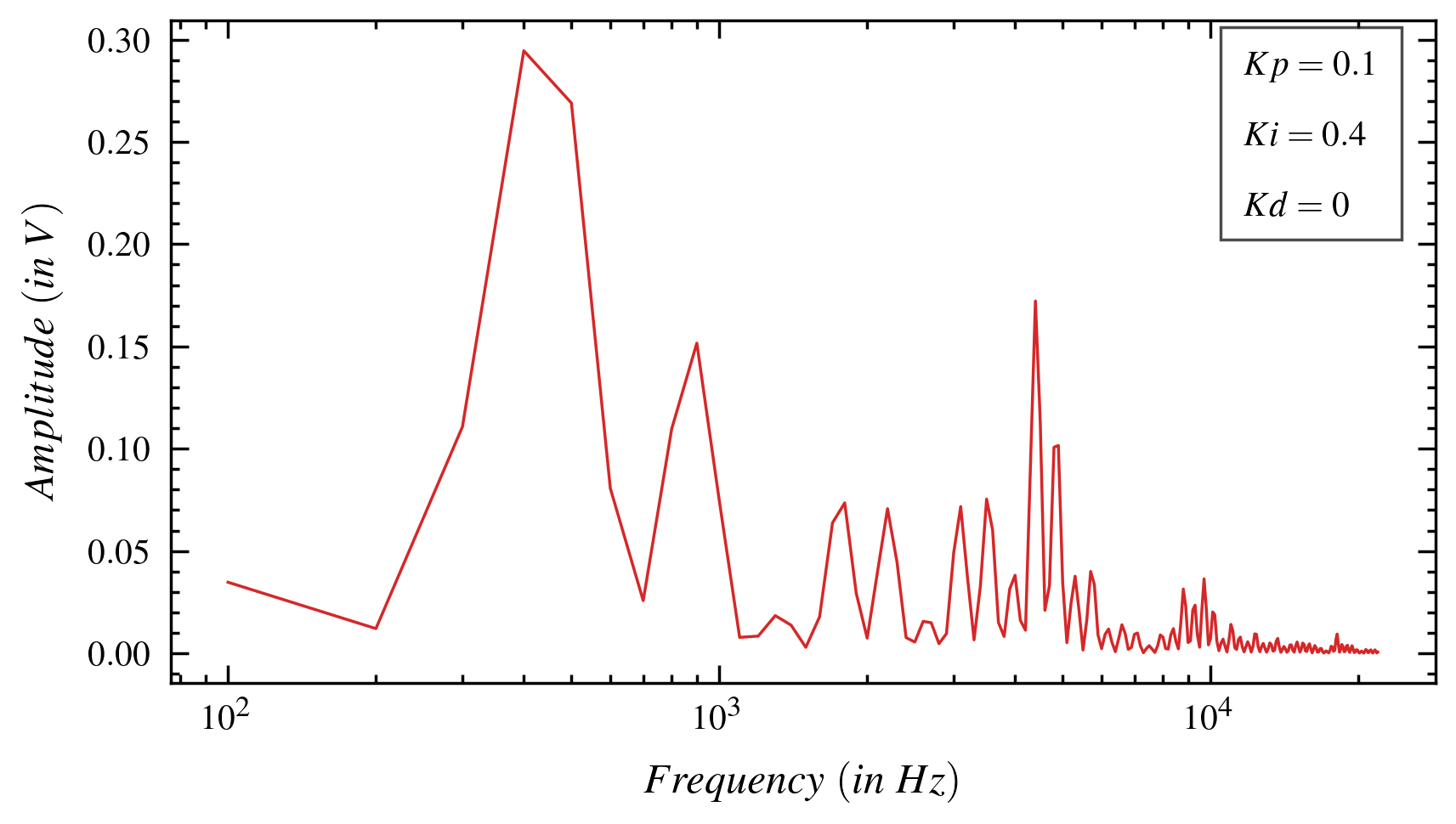}\label{F7a}}
\subfloat[2x Oversampling]{\includegraphics[width=0.24\textwidth]{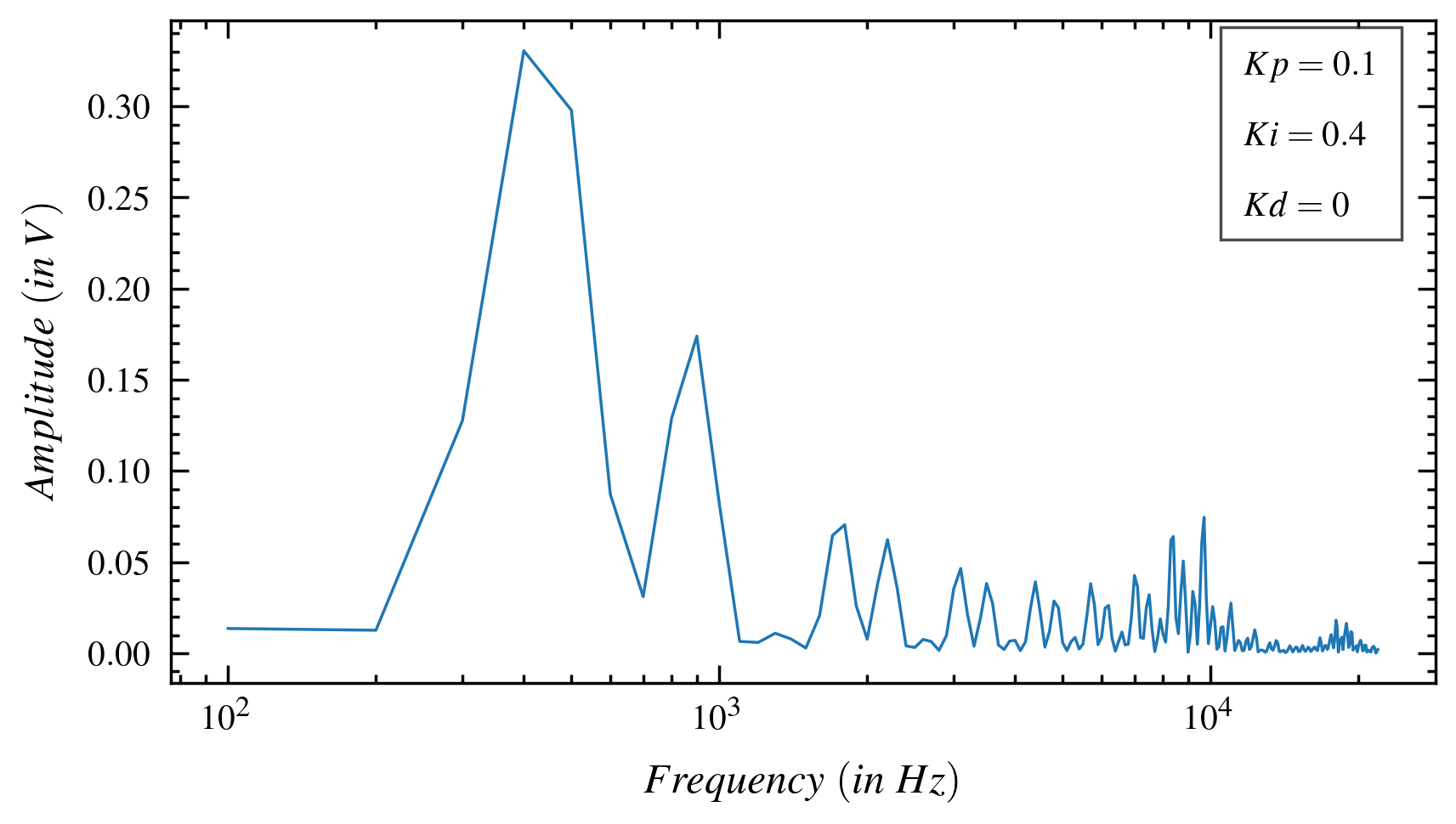}\label{F7b}}\hfill
\subfloat[4x Oversampling]{\includegraphics[width=0.24\textwidth]{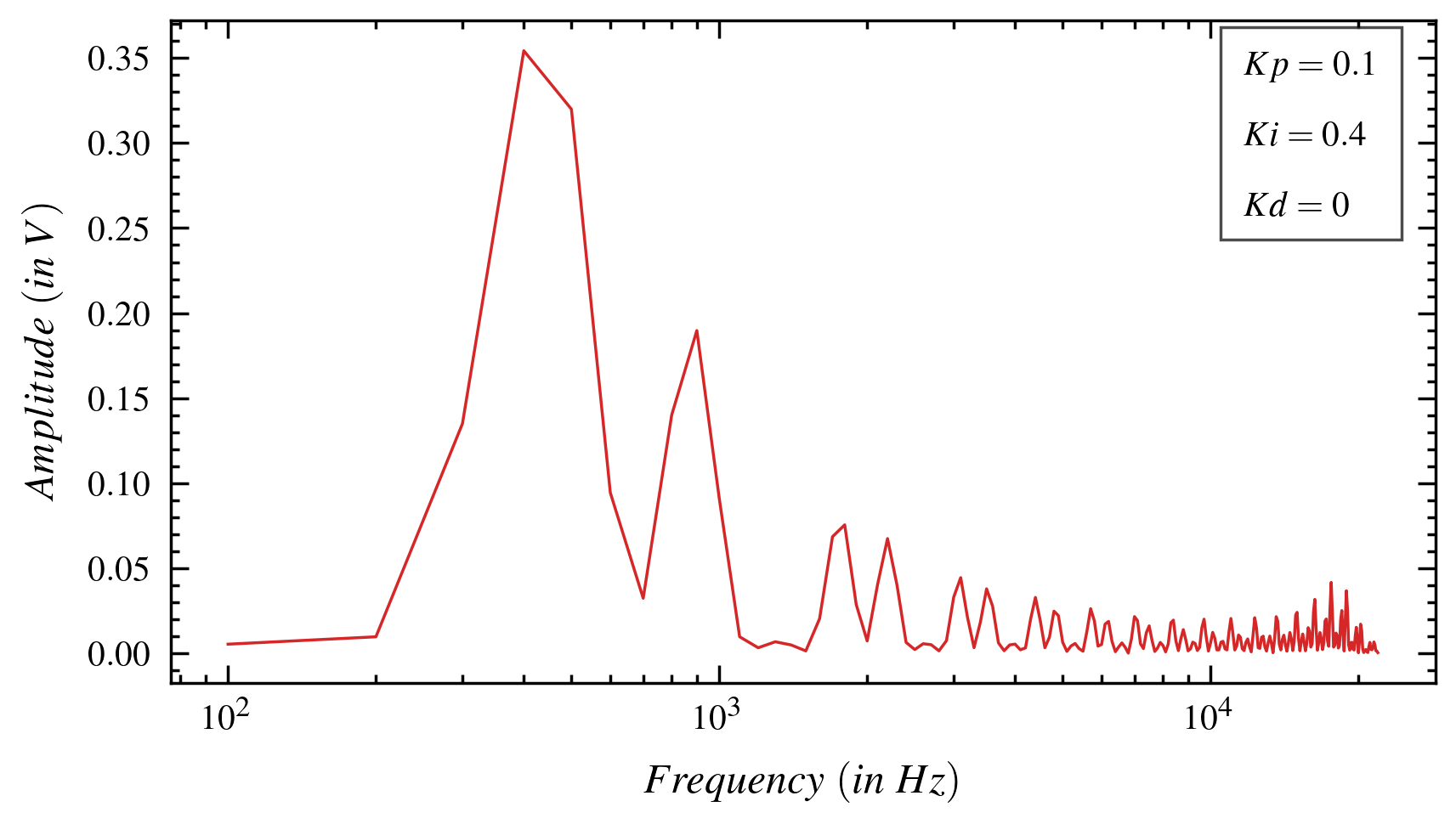}\label{F7c}}
\subfloat[8x Oversampling]{\includegraphics[width=0.24\textwidth]{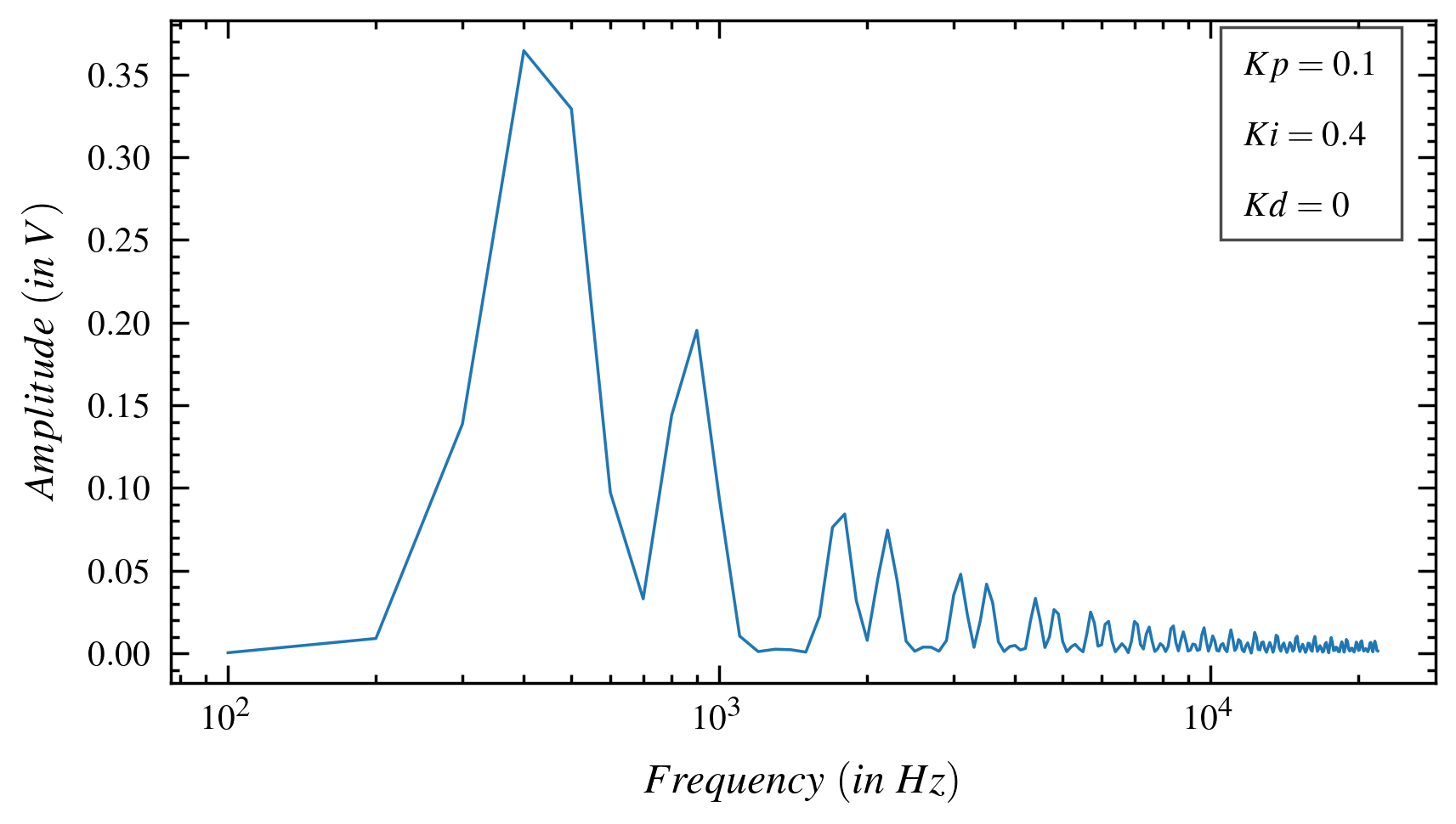}\label{F7d}}\hfill
\caption{As the magnitude of oversampling increases, aliased components in the PIDS $output$ decreases. However, higher degrees of oversampling requires considerable processing power and may not be feasible on low powered processors.}
\label{F7}
\end{figure}

As part of this research, numerous methods were prototyped and integrated into the PIDS framework to prevent aliasing intrinsically. However, most of them returned insufficient degrees of success. Finally, oversampling was employed as the means of anti-aliasing for PIDS synthesized signals. Initially, the $output$ is generated at a sampling rate of  2-, 4- or 8- times the playback rate. Next, the $output$ was convoluted with a windowed, low-pass FIR filter having a cutoff frequency of 20 kHz and sufficient transition bandwidth to truncate all the high-frequency components susceptible to aliasing. Finally, the filtered $output$ was downsampled back to the playback rate. A comparison of the $output$ produced in various scenarios of oversampling is depicted in \cref{F7}. While higher degrees of oversampling reduce the occurrences of rolled over components appearing in the final synthesized $output$, they may increase computational resource utilization and, therefore, may not be feasible with low powered processors.  In such cases, implementers must arrive at a suitable tradeoff between audio quality and processing costs.

\section{Effect of PIDS Controls}
\label{S4}

Similar to PID control, the $output$ synthesized by PIDS largely depends on the values of $K_{p}$, $K_{i}$ and $K_{d}$ provided as input by the user. To understand their working and effect on the $output$ produced, a software version of PIDS is implemented in Python3. In each case (unless specified), the $output$ is generated against a step $artist$ of 440 Hz. The $artist$ is portrayed by \cref{F4a}, and its breakpoints are documented by \cref{T1}.

\begin{table}[h!]
\centering
\begin{tabular}{|c | c | c |}
 \hline
 Breakpoint Number & X-Value & Y-Value \\ 
 \hline
 1 & 0 & 0 \\
 2 & 0.33 & 1 \\
 3 & 0.67 & -1 \\
 4 & 1 & 0 \\ [1ex] 
 \hline
\end{tabular}
\caption{Set of breakpoints used for studying the effects of various PIDS inputs}
\label{T1}
\end{table}

\subsection{Proportional Gain ($K_{p}$)}
\label{S41}

\begin{figure}[t]
\centering
\subfloat[]{\includegraphics[width=0.24\textwidth]{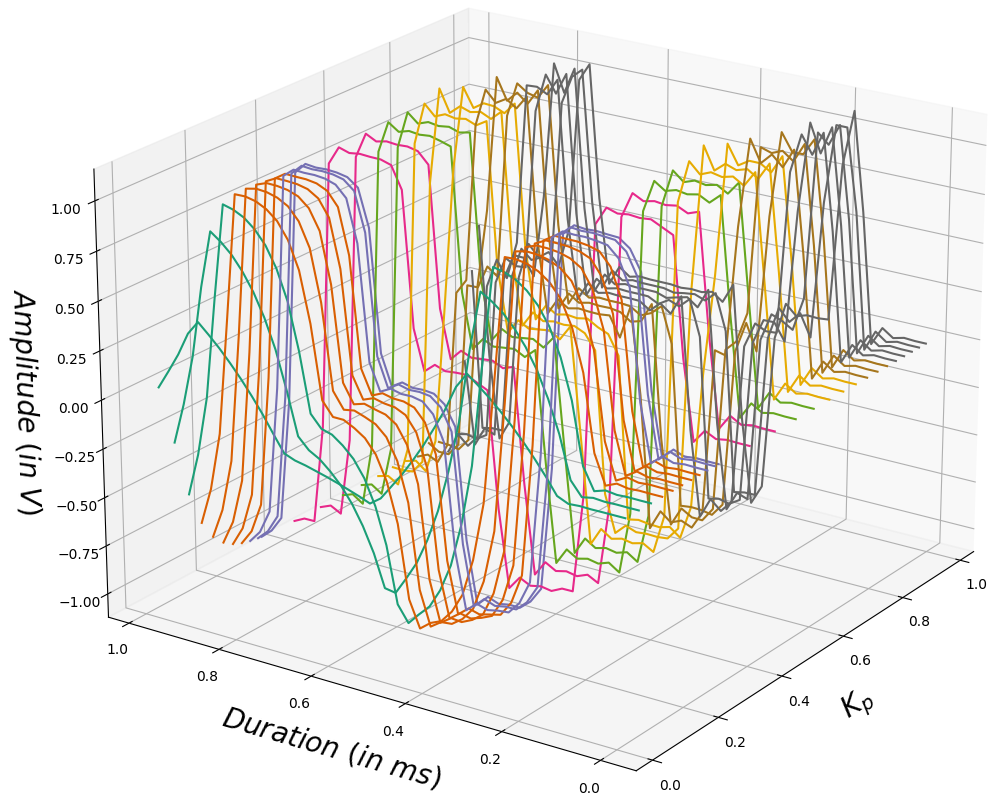}\label{F9a}}
\subfloat[]{\includegraphics[width=0.24\textwidth]{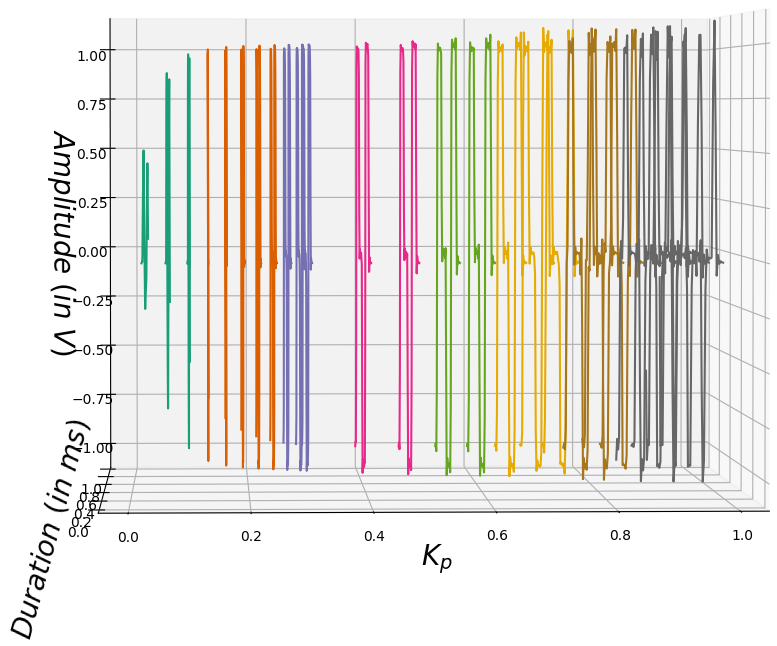}\label{F9b}}
\caption{Time-domain plot of the effect of varying $K_{p}$ on PIDS $output$.}
\label{F9}
\end{figure}

\begin{figure}[t]
\centering
\subfloat[]{\includegraphics[width=0.24\textwidth]{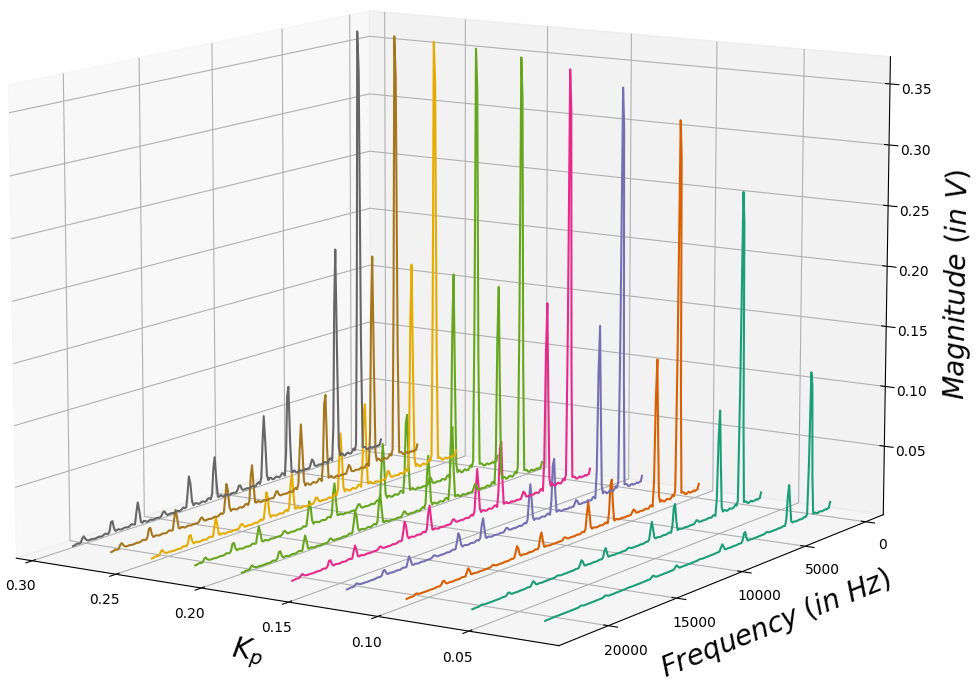}\label{F10a}}
\subfloat[]{\includegraphics[width=0.24\textwidth]{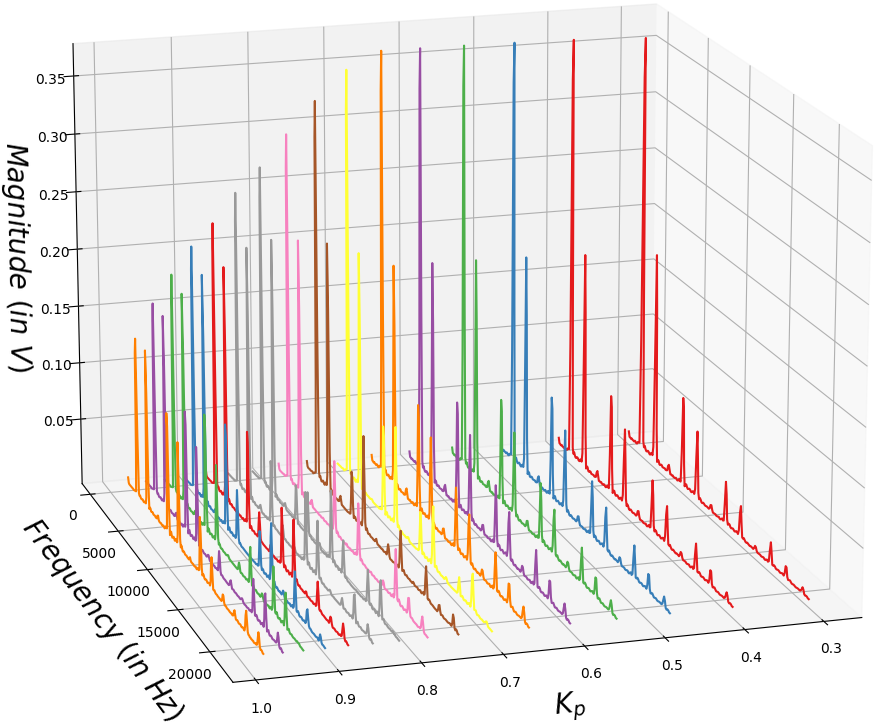}\label{F10b}}
\caption{Frequency-domain plot of the effect of varying $K_{p}$ on PIDS $output$.}
\label{F10}
\end{figure}

The $K_{p}$ parameter controls the amount of contribution the P-component of PID control has on the synthesized $output$. The P-component is a function of the error (i.e., difference) between instantaneous SP and PV. When the system is stable, higher values of $K_{p}$ ensure that this error is reduced more quickly. For PIDS, as $K_{p}$ increases, the $output$ follows the $artist$ more closely (the phase-lag between them reduces). Additionally, by controlling the extent of this lag, the $K_{p}$ parameter indirectly determines the amplitude of the synthesized $output$. As $artists$ values change considerably at breakpoints, the lag present may prevent the $output$ from reaching the breakpoint y-values at lower values of $K_{p}$. \Cref{F9a} illustrates the change of waveforms as $K_{p}$ increases from 0 to 1;  the $output$ increasingly resembles the $artist$.

Moreover, \cref{F9b} shows the increase in the $output$ amplitude on increasing $K_{p}$, as it can follow the $artist$ better. \Cref{F10} depicts the frequency spectrum of \cref{F9}. At lower values of $K_{p}$, the frequency distribution remains almost constant as the energy of each frequency component increases per the increasing amplitude. However, at higher $K_{p}$, the $output$ starts resembling the step-$artist$. Consequently, while the total energy remains the same, it gets more evenly distributed among the immediate harmonics (step $artists$ have infinite odd harmonics).

\subsection{Integral Gain ($K_{i}$)}
\label{S42}

\begin{figure}[t]
\centering
\includegraphics[width=0.48\textwidth]{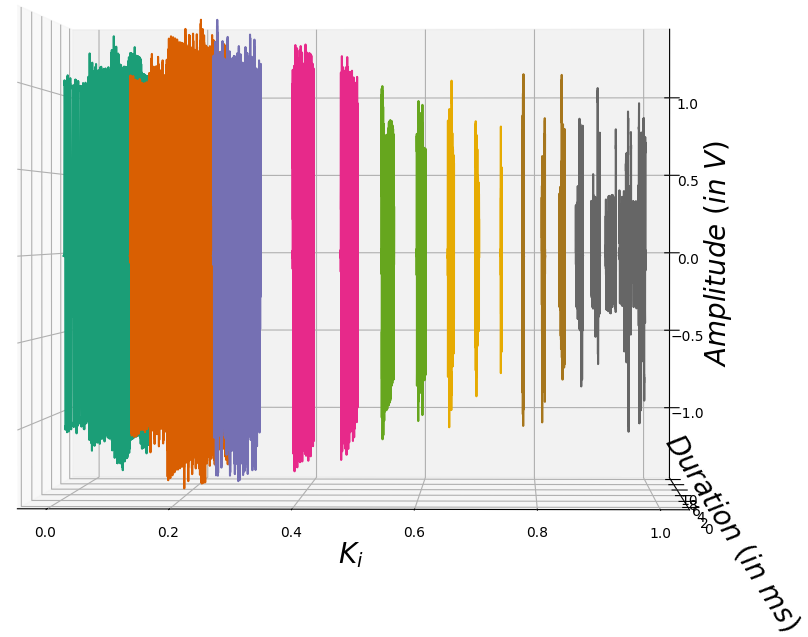}
\caption{Time-domain plot of the effect of varying $K_{i}$ on PIDS $output$.}
\label{F11}
\end{figure}

\begin{figure}[b]
\centering
\subfloat[]{\includegraphics[width=0.24\textwidth]{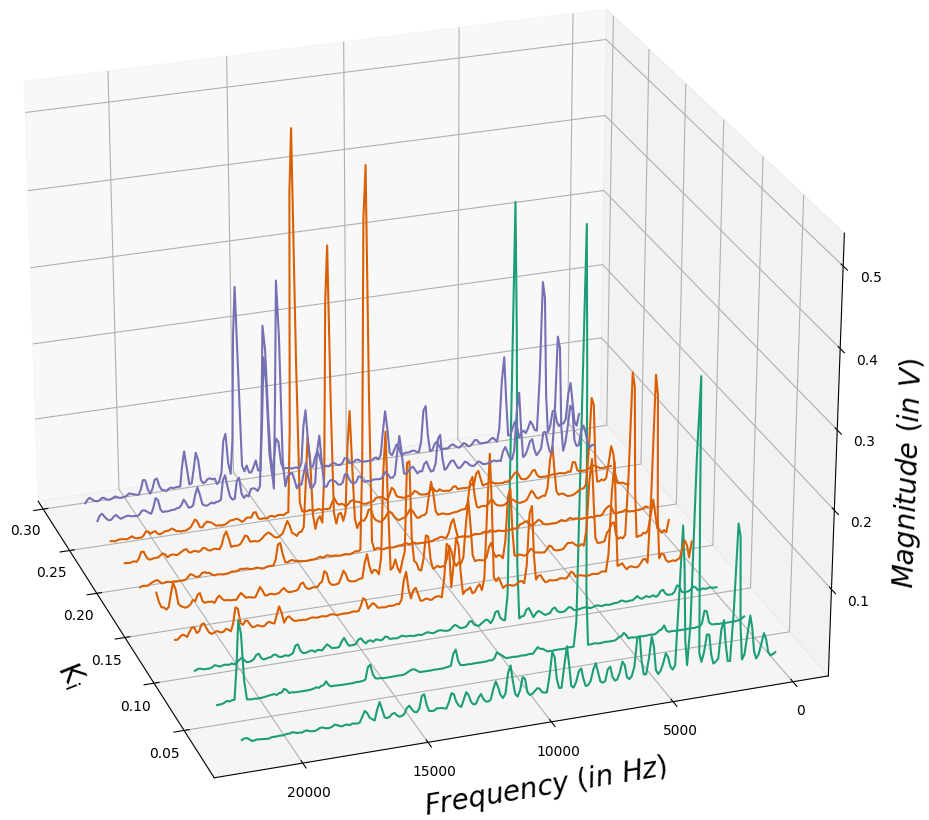}\label{F12a}}
\subfloat[]{\includegraphics[width=0.24\textwidth]{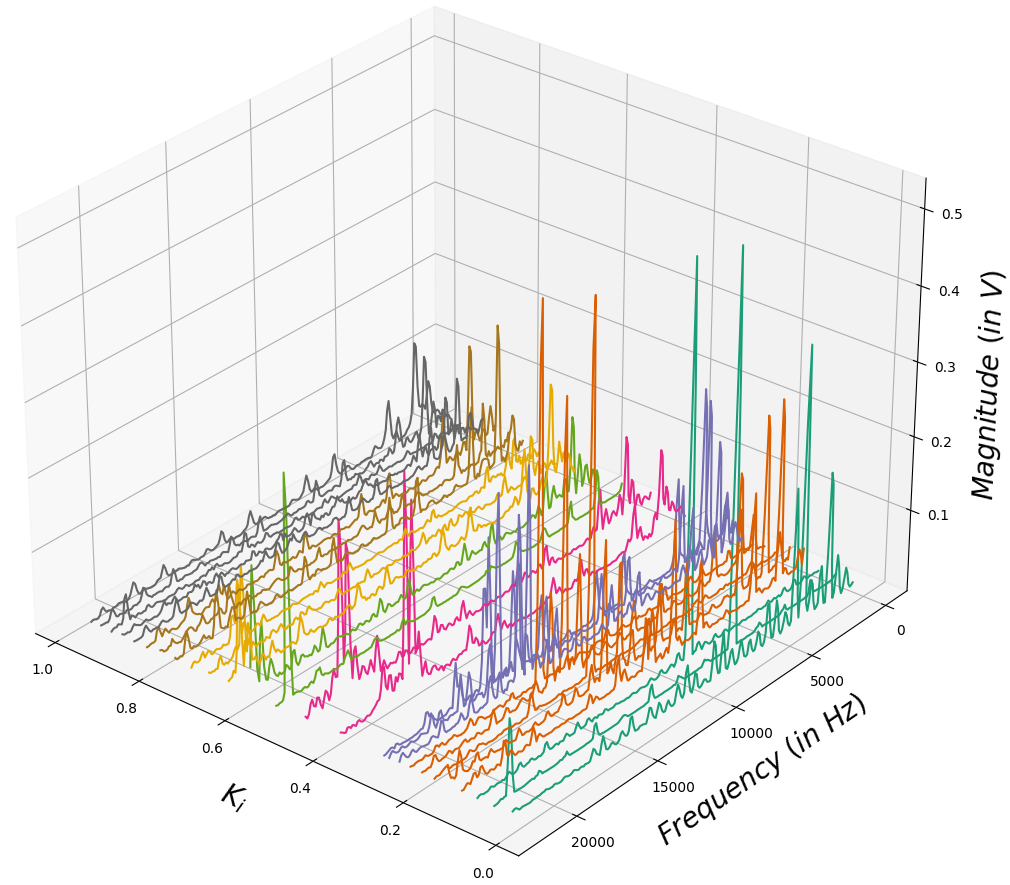}\label{F12b}}
\caption{Frequency-domain plot of the effect of varying $K_{i}$ on PIDS $output$.}
\label{F12}
\end{figure}

The $K_{i}$ parameter controls the extent to which the I-component of PID control influences the synthesized $output$. Unlike the P-component, the historical values of errors are considered here, and corrective action is produced as a function of the error accumulated over time. The impact of accumulation is that the PV will eventually reach the SP as the integral (accumulated error) decays back to 0. In PIDS, this results in two observations.

Firstly, activating the I-components will ensure that the $output$ can reach the $artist$ breakpoints at lower values of $K_{i}$ than $K_{p}$. This can be deduced by comparing the teal waveforms of \cref{F9b} and \cref{F11}.

Secondly, the process of integral decay and consequent accumulation in the opposite direction introduces high-frequency oscillations into the $output$. As evidenced by \cref{F12a}, the higher the value of $K_{i}$, the higher the resulting $output$'s peak frequency. From a music producer's perspective, noting this phenomenon is important as the pitch of the synthesized audio signal may be drastically higher than expected when playing a note. To prevent such cases, an alternative would be to add in a small amount of the P-component while using the I-component, thereby operating PIDS in the PI-mode. Moreover, as $K_{i}$ increases, the induced harmonics can exceed the Nyquist frequency, as evidenced for waveforms with $K_{i}$ greater than 0.6 in \cref{F12b}. In these cases, the anti-aliasing mechanism kicks in and removes considerable energy from the synthesized $output$.  Consequently, the average magnitude of such signals is lesser than those with lower $K_{i}$, as shown in \cref{F11}.

\subsection{Derivative Gain ($K_{d}$)}
\label{S43}

\begin{figure}[!b]
\centering
\subfloat[]{\includegraphics[width=0.24\textwidth]{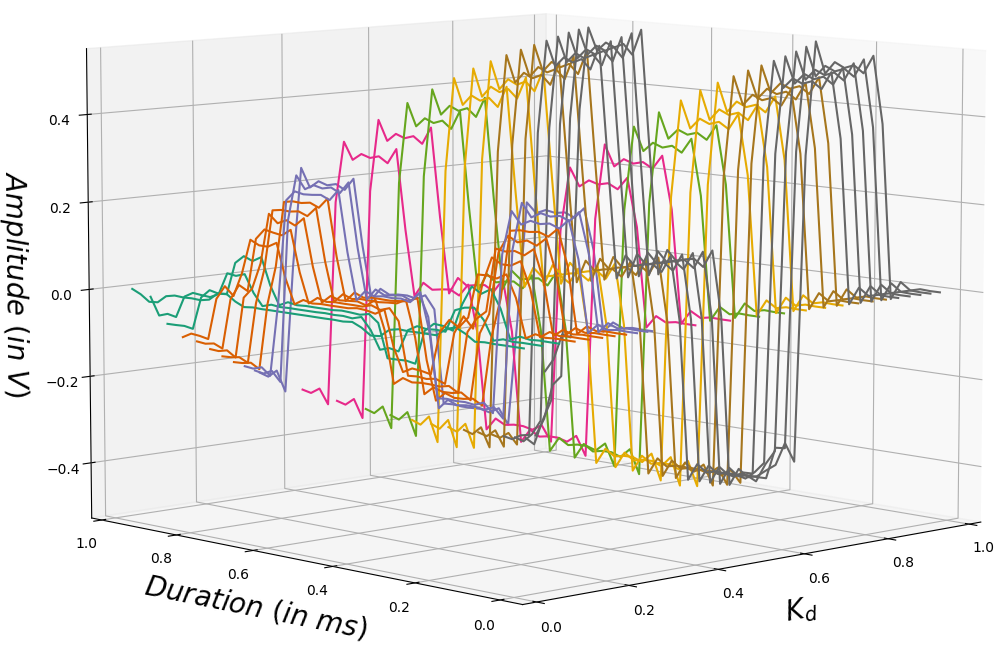}\label{F13a}}
\subfloat[]{\includegraphics[width=0.24\textwidth]{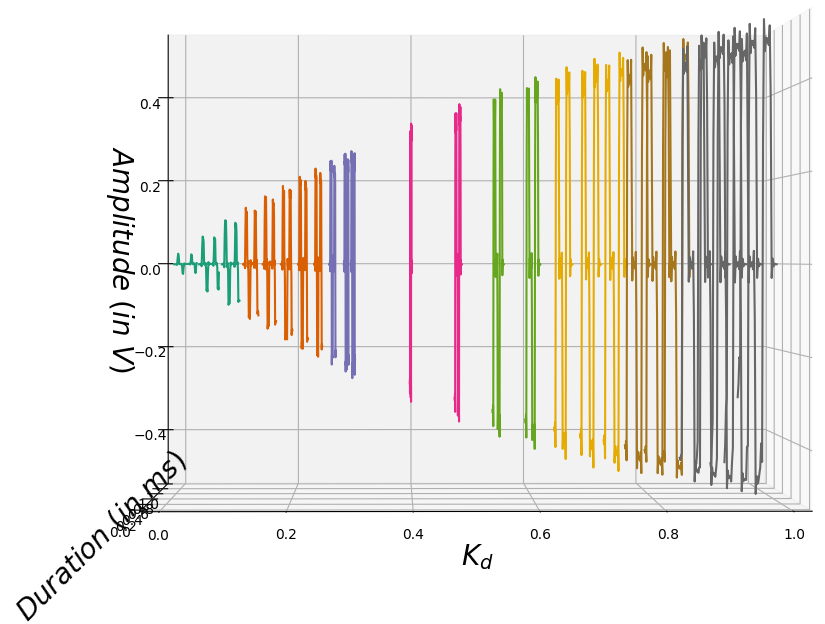}\label{F13b}}
\caption{Time-domain plot of the effect of varying $K_{d}$ on PIDS $output$.}
\label{F13}
\end{figure}

\begin{figure}[!b]
\centering
\includegraphics[width=0.48\textwidth]{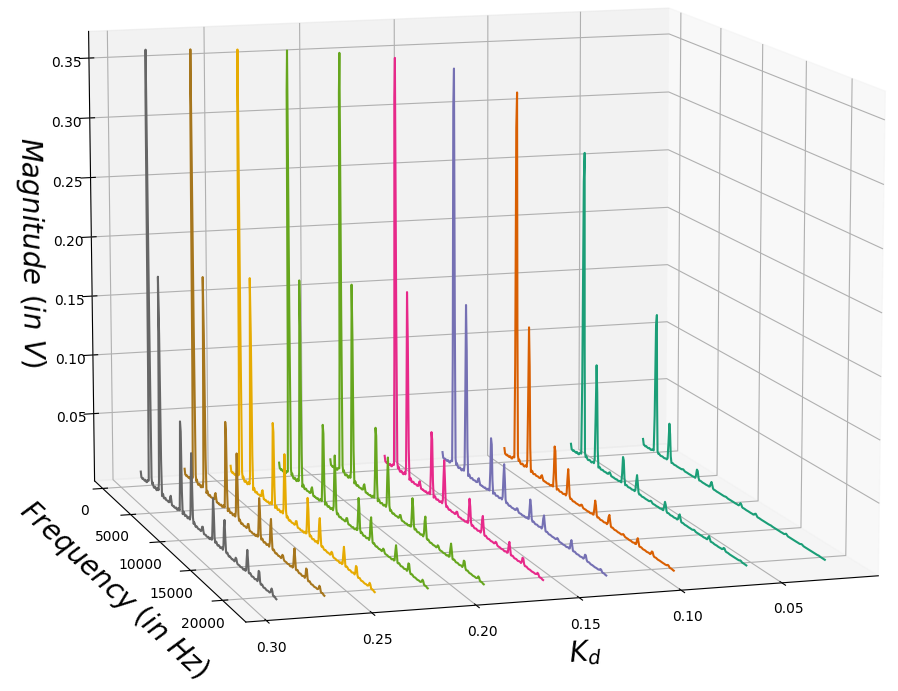}
\caption{Frequency-domain plot of the effect of varying $K_{d}$ on PIDS $output$.}
\label{F14}
\end{figure}

The $K_{d}$ parameter controls the presence of the D-component of PID control in the synthesized $output$. In conventional controllers, the D-component acts on the derivative of errors to provide a lead compensation to account for the sluggish performance of larger process systems (that act as high-capacity integrators). In PIDS, the D-component is simply a function of the difference between consecutive errors. Consequently, when only $K_{d}$ is active and is less than one (negative feedback), the $output$ of PIDS is a miniaturized form of the $artist$, the amplitude of the $output$ being proportional to the value of $K_{d}$. This can be observed in \cref{F13}. However, at higher values, the system tends to become unstable (due to positive feedback), and the waveforms thus generated may not yield desirable audio signals.

Analyzing the frequency spectrum of purely D-mode waveforms provided by \cref{F14}, it is clear that increasing the value of $K_{d}$ does not modify the frequency distribution of the $output$ (it shares the same spectrum as the $artist$). Overall, the $K_{d}$ parameter is more suitable to be used while the $K_{i}$ parameter is also activated. In such cases, the $K_{d}$ tends to pull back the effect of $K_{i}$ by providing a reduced form of P-component contribution.

\subsection{Multi-modes}
\label{S44}

While PIDS can be operated to synthesize audio signals with just one of $K_{p}$, $K_{i}$ and $K_{d}$ being activated, its flexibility in terms of controlling the waveforms produced lies in using it with multiple parameters activated. This includes all the permutations of PID control: PI, PD, ID and PID-modes. The behaviour of PIDS in each of these multi-modes is the combined effect of the activated parameters, the individual contribution of each parameter to the $output$ being proportional to the values of their relative gains.

\begin{figure}[!b]
\centering
\includegraphics[width=0.48\textwidth]{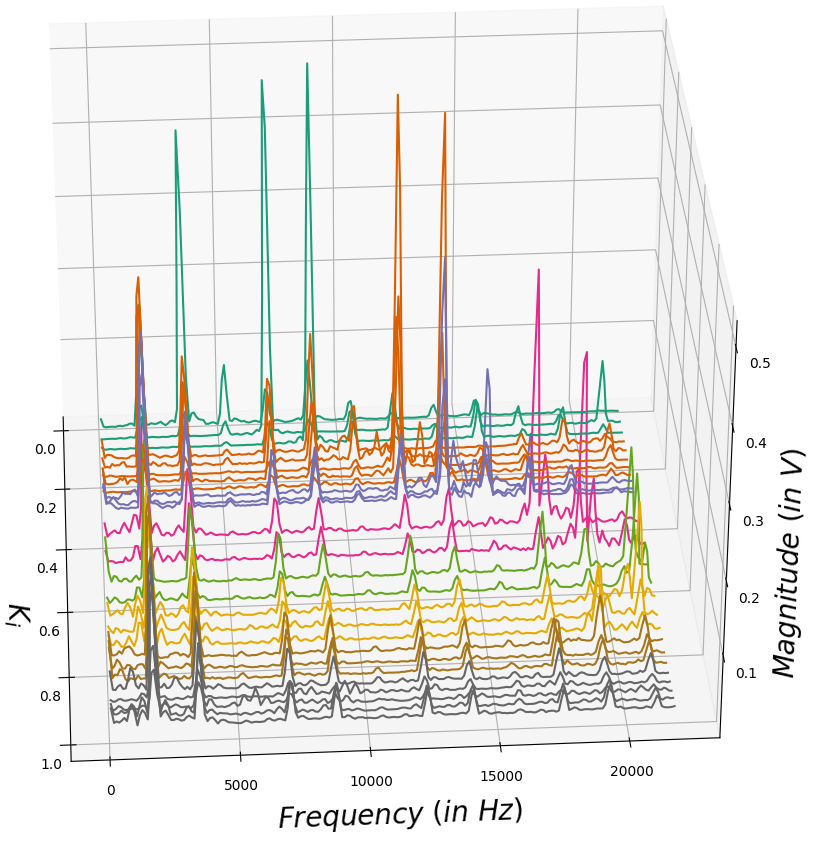}
\caption{Frequency-domain plot of PIDS $outputs$ in the PI mode.}
\label{F15}
\end{figure}

\Cref{F15} illustrates the frequency distribution of the $outputs$ produced in the PI mode, with the $K_{p}$ remaining constant at 0.6 while $K_{i}$ is sweeped from zero to one. The resulting $outputs$ not only have considerable energy at the fundamental frequency of the $artist$ (440 Hz) that is induced by the P-component but also have a significant high frequency component introduced by the I-component.

\subsection{Breakpoints}
\label{S45}

\begin{figure}[t]
\centering
\subfloat[PIDS $output$ for the triangular $artist$]{\includegraphics[width=0.24\textwidth]{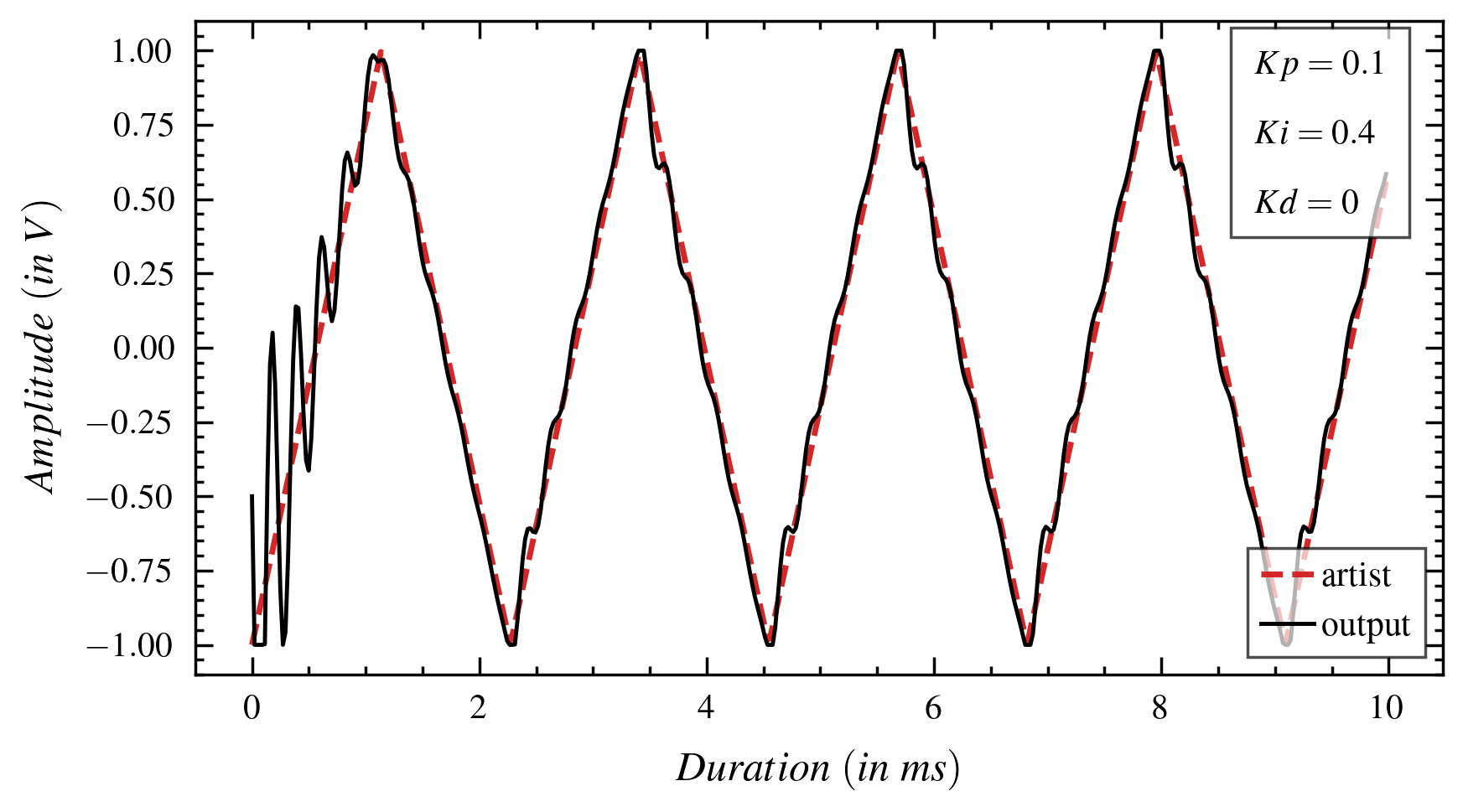}\label{F24a}}
\subfloat[PIDS $output$ for the sawtooth $artist$]{\includegraphics[width=0.24\textwidth]{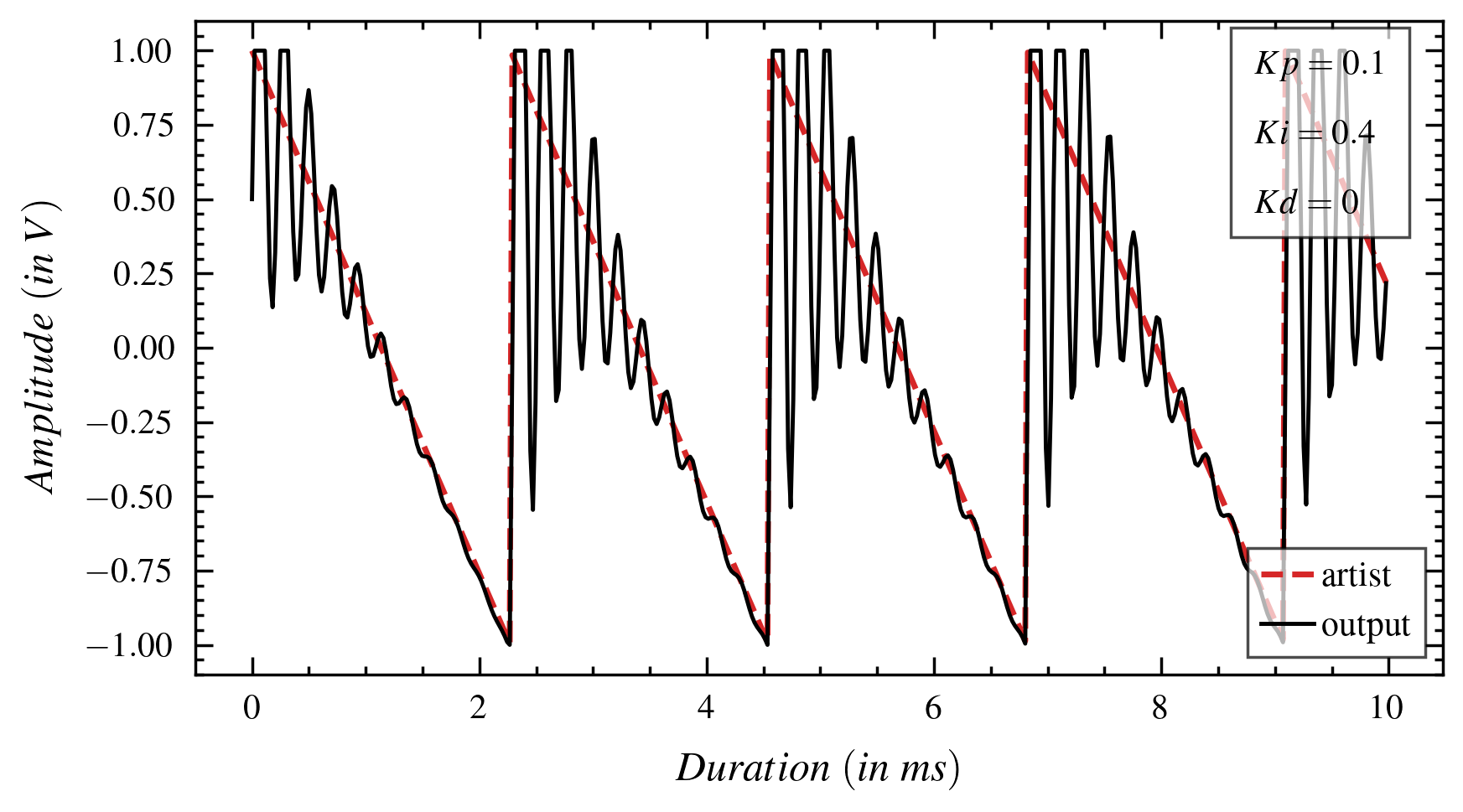}\label{F24b}}
\caption{The placement of $artist$ breakpoint coordinates determines the $artist$ shape and hence, the resulting PIDS $output$. As seen in the two plots, even for the same $artist$ type (linear), varying a single breakpoint x-value can lead to a considerable change in the $output$.
}
\label{F24}
\end{figure}

As described in \cref{S2}, the set of $artist$ breakpoint coordinates determines its subtleties and, therefore, the resulting $output$ synthesized by PIDS. The user can control the y-values of the first and last breakpoints and the x- and y-values of the intermediate breakpoints. By modulating these values with time, a great degree of variation is achieved in the type of $artists$ obtained. Consequently, the audio signal generated in this process may vary drastically as the modulation sweeps, producing an effect similar to wavetable synthesis discussed in detail in \cref{S64}. \Cref{F24} describes the changes in PIDS $artists$ (and hence, the $output$) as the breakpoints change for a linear $artist$.

\section{Effect of Fundamental Artist Frequency}
\label{S5}

\begin{figure}[b]
\centering
\subfloat[]{\includegraphics[width=0.24\textwidth]{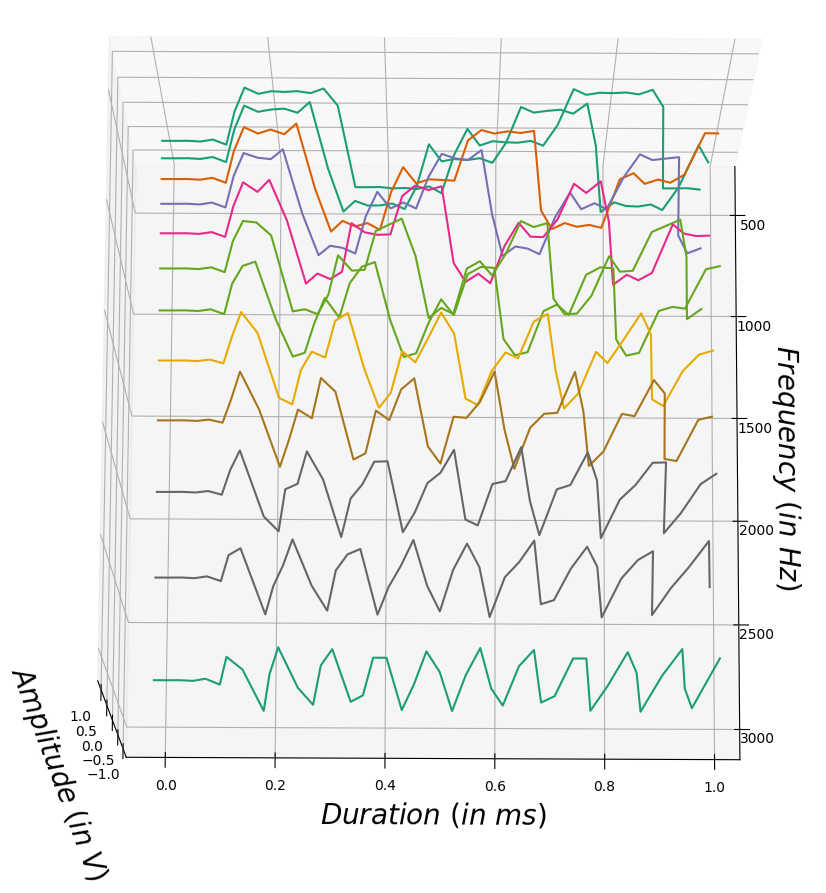}\label{F16a}}
\subfloat[]{\includegraphics[width=0.24\textwidth]{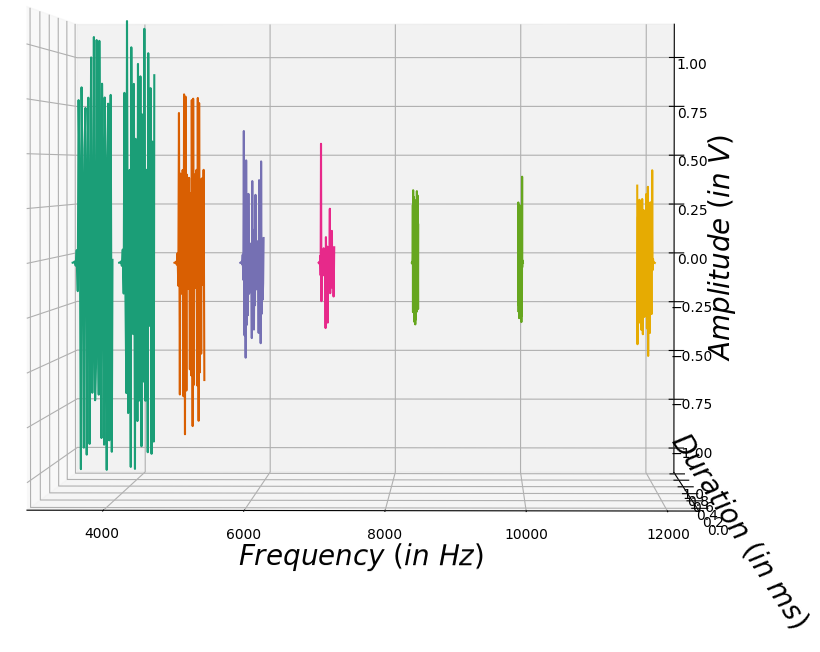}\label{F16b}}
\caption{Time-domain plot for $K_{p}$ = 0.6 and $K_{i}$ = 0.3 at various fundamental $artist$ frequencies. The fundamental $artist$ frequency influences the fundamental frequency of the PIDS $output$. The $output$ can trace the $artist$ more faithfully at lower values, and these frequencies are the same. As $artist$ frequency increases, the $output$ loses intricate details resulting in a visible change in $output$ waveforms. Anti-aliasing measures, if included, can also result in a decrease in $output$ amplitude due to removal of higher harmonics.}
\label{F16}
\end{figure}

\begin{figure}
\centering
\includegraphics[width=0.48\textwidth]{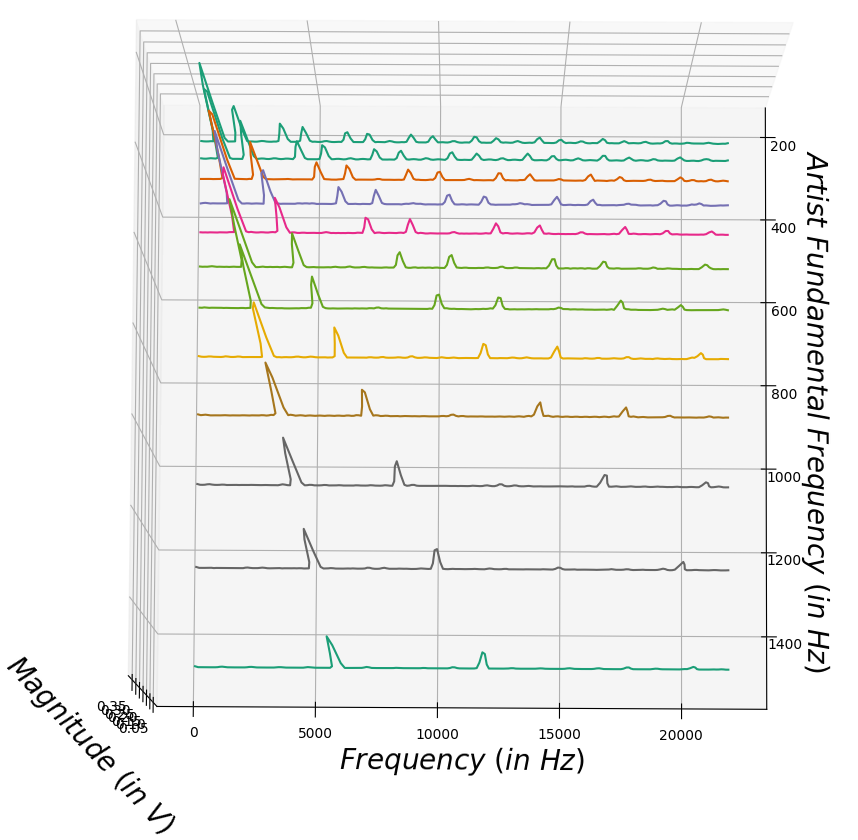}
\caption{Frequency-domain plot for \cref{F16}.}
\label{F17}
\end{figure}

The fundamental frequency of PIDS $output$ is dependent mainly on the fundamental frequency of the $artist$. At lower frequencies, the $output$ can follow the $artist$ more faithfully, and hence its peak frequency accurately matches the fundamental frequency. This can be observed in \cref{F16a} and \cref{F17}, where the $outputs$ were generated against a step $artist$ having breakpoints given by \cref{T1}. However, as this frequency increases for higher notes, the finer details of the synthesized waveforms get lost as the $output$ struggles to match the faster oscillations of the $artist$. Consequently, for the same input provided to PIDS, the timbre generated at lower octaves may vary noticeably than at higher octaves. Another effect observed in this case is that the waveforms' amplitude also tends to fall as the note frequency increases, the reason being the loss of finer details and the band limit set by the anti-aliasing (removing most of the harmonics). This phenomenon is observed in \cref{F16b} and occurs even for conventional sine, triangle and sawtooth oscillators. Additionally, such situations arise only for the notes belonging to the higher octaves. For the commonly used octaves of modern music, the PIDS $output$ generated is majorly consistent and has the fundamental frequency characteristic of the note played.

\section{Applications}
\label{S6}

\subsection{Additive Synthesis}
\label{S61}

\begin{figure}[t]
\centering
\includegraphics[width=0.48\textwidth]{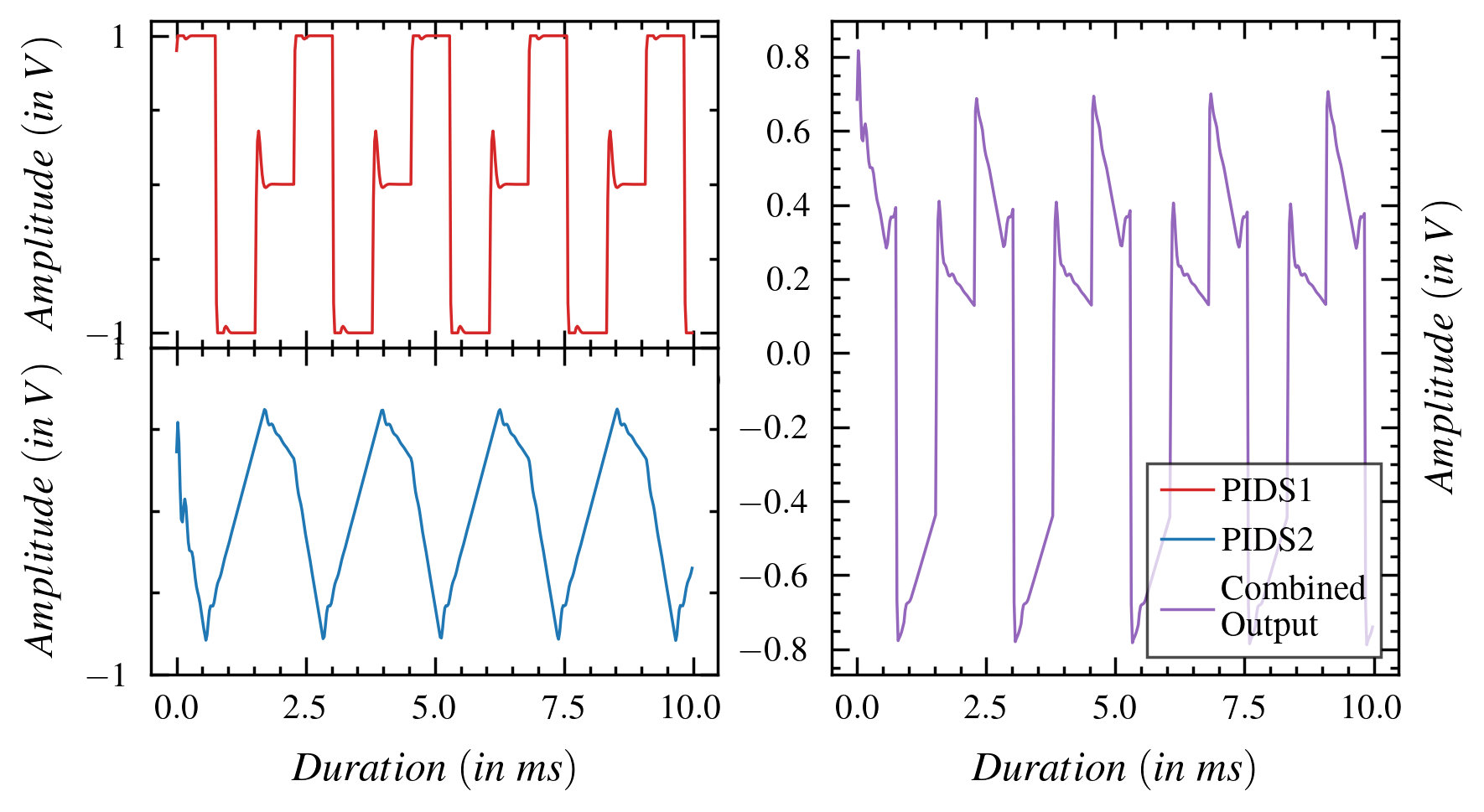}
\caption{$outputs$ from multiple PIDS can be combined using conventional additive synthesis techniques. The resulting $output$ combines the features of its constituents in proportion to the ratios in which they are fused, potentially resulting in complex waveforms.}
\label{F18}
\end{figure}

While a single instance of PIDS can generate a wide variety of waveforms, greater degrees of freedom are provided to the user by combining two or more PIDS $outputs$. In such cases of additive synthesis, two more user-controls may be provided: a ratio adjuster to set the proportion by which the individual PIDS $outputs$ are mixed to form the final $output$, and a detuner to transpose one of the PIDS with respect to the other. Additionally, if the set of breakpoints y-values of all the PIDS is the same, the relative difference in their x-values can implement phase differences between them. \Cref{F18} depicts a simple example of additive synthesis, with one PIDS having a step $artist$ while the other having a linear $artist$.

\subsection{Low Frequency Oscillator (LFO)}
\label{S62}

When used with $artist$ frequencies under 20 Hz, PIDS effectively behaves as an LFO. Hence, its $output$ can be used to modulate parameters like the pitch, gain and cutoff frequencies of other audio modules (including conventional synthesizers and effect processors). In PIDS LFOs, the user controls and underlying processing remain the same as for audio synthesis, the only crucial difference being an optional lack of anti-aliasing in LFOs. Aliasing may be allowed for two main reasons; firstly, the harmonics generated for LFOs over the Nyquist frequency are significantly attenuated. Hence, the effect of any folding taking place is mostly inaudible. Secondly, implementing anti-aliasing in LFO $output$ may not necessarily guarantee the absence of aliased components in the signal it is modulating (therefore, anti-aliasing must be performed directly on the modulated signal).

\begin{figure}[b]
\centering
\subfloat[PIDS LFO $output$ in time domain.]{\includegraphics[width=0.24\textwidth]{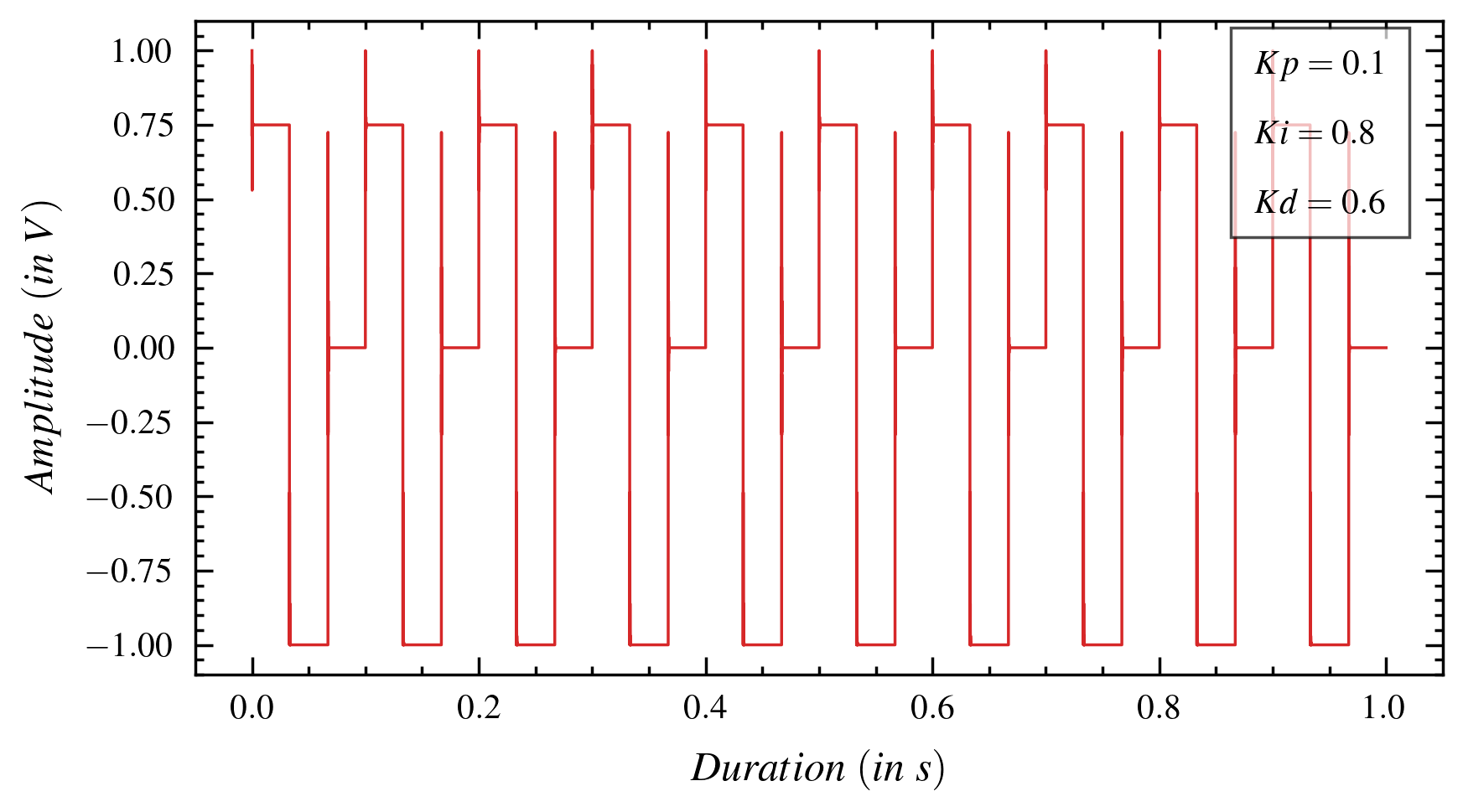}\label{F19a}}
\subfloat[PIDS LFO $output$ in frequency domain.]{\includegraphics[width=0.24\textwidth]{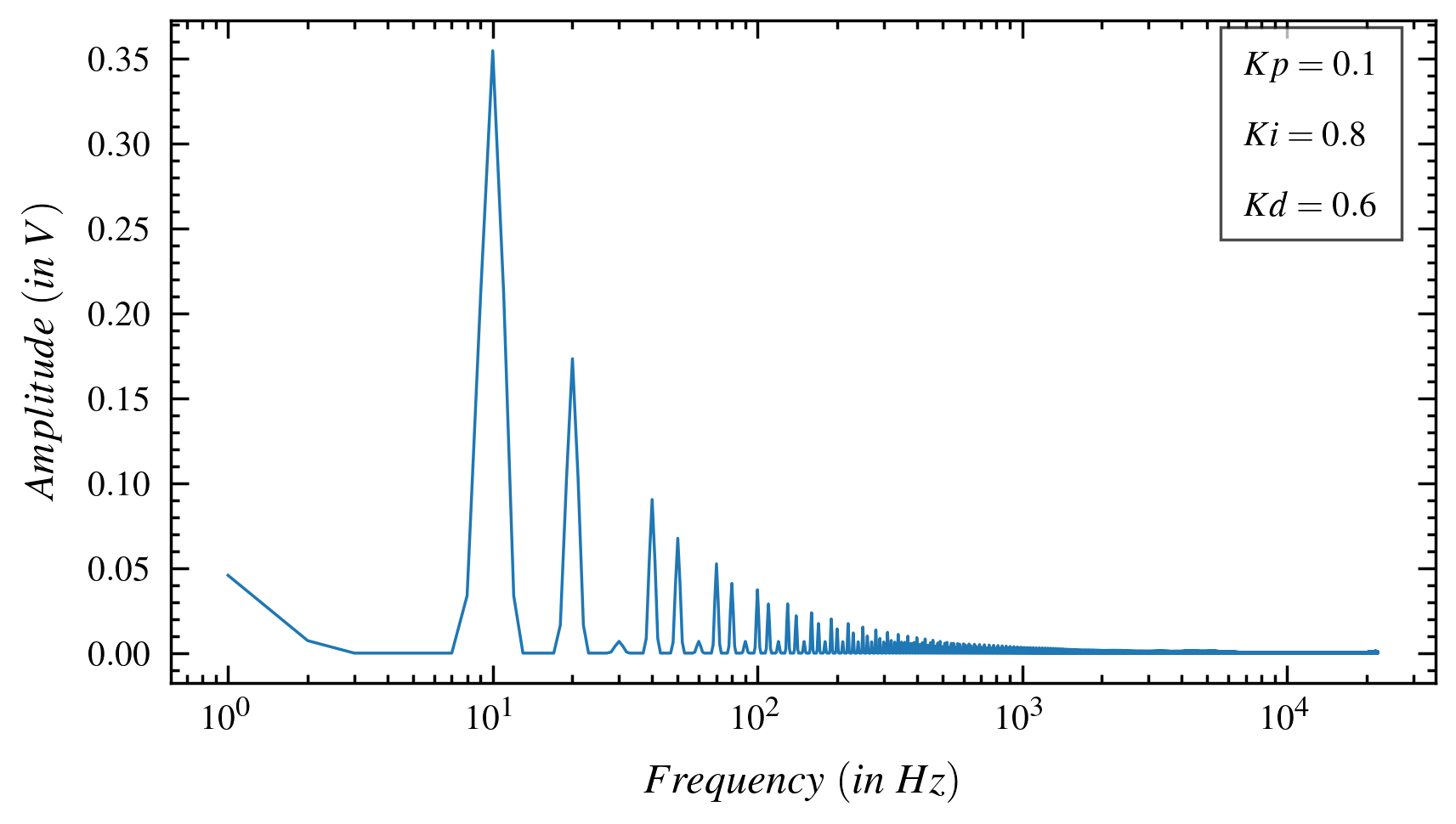}\label{F19b}}
\caption{Using $artist$ frequencies less than 20 Hz, PIDS can be used to generate LFOs. PIDS controls can be modulated to shape LFOs in a manner exactly similar to generating audio signals.}
\label{F19}
\end{figure}

On the other hand, it must be noted that the LFOs generated with the $K_{i}$ dominating may generate high-frequency harmonics that get rolled over. In such situations, the implementers may choose to enable anti-aliasing as soon as the I-component dominates (if such foldovers produce undesirable effects). An example of PIDS generated LFO is demonstrated by \cref{F19}.

\subsection{Alternative to FM Synthesis}
\label{S63}

\begin{figure}[t]
\centering
\subfloat[Time-domain]{\includegraphics[width=0.24\textwidth]{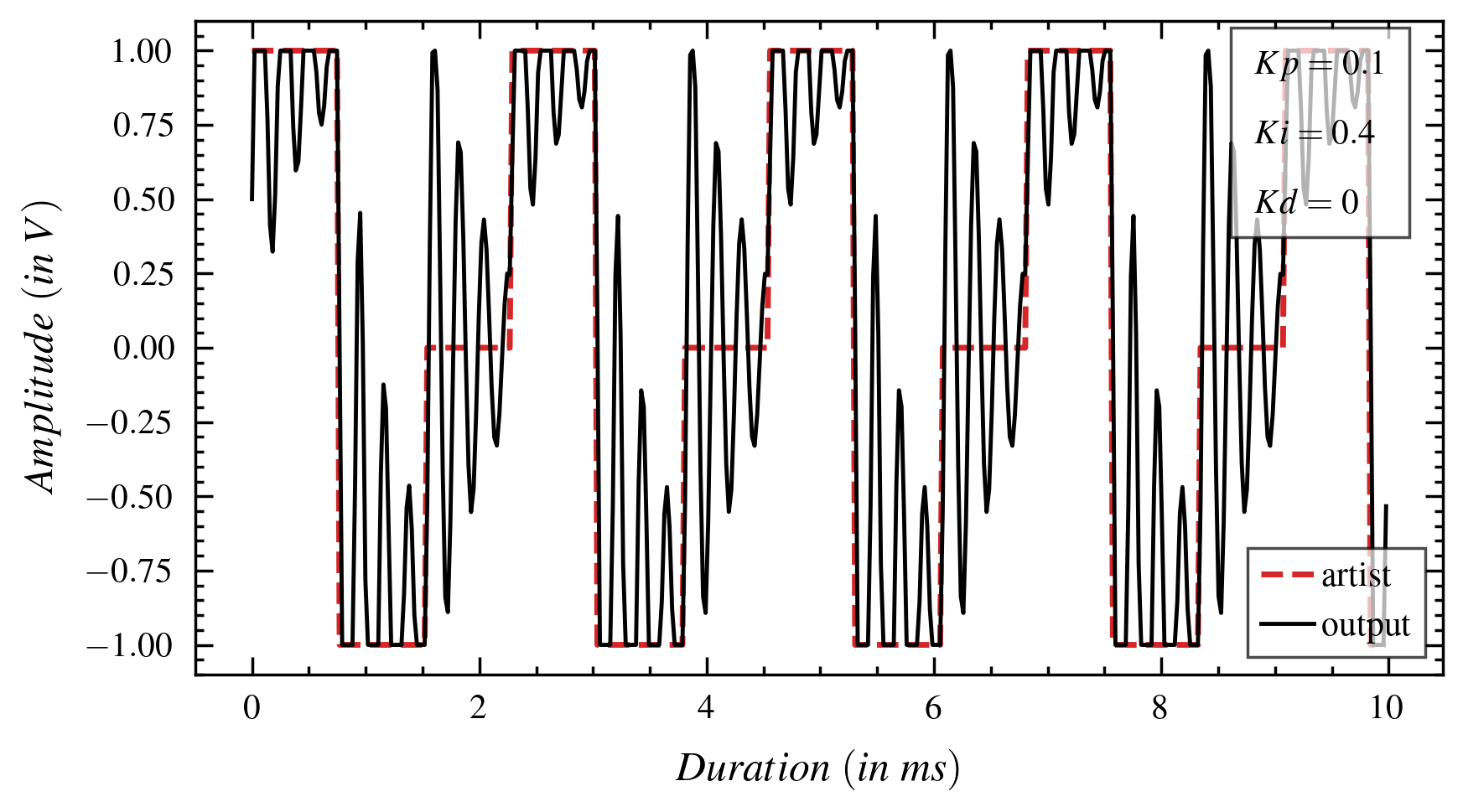}\label{F23a}}
\subfloat[Frequency-domain]{\includegraphics[width=0.24\textwidth]{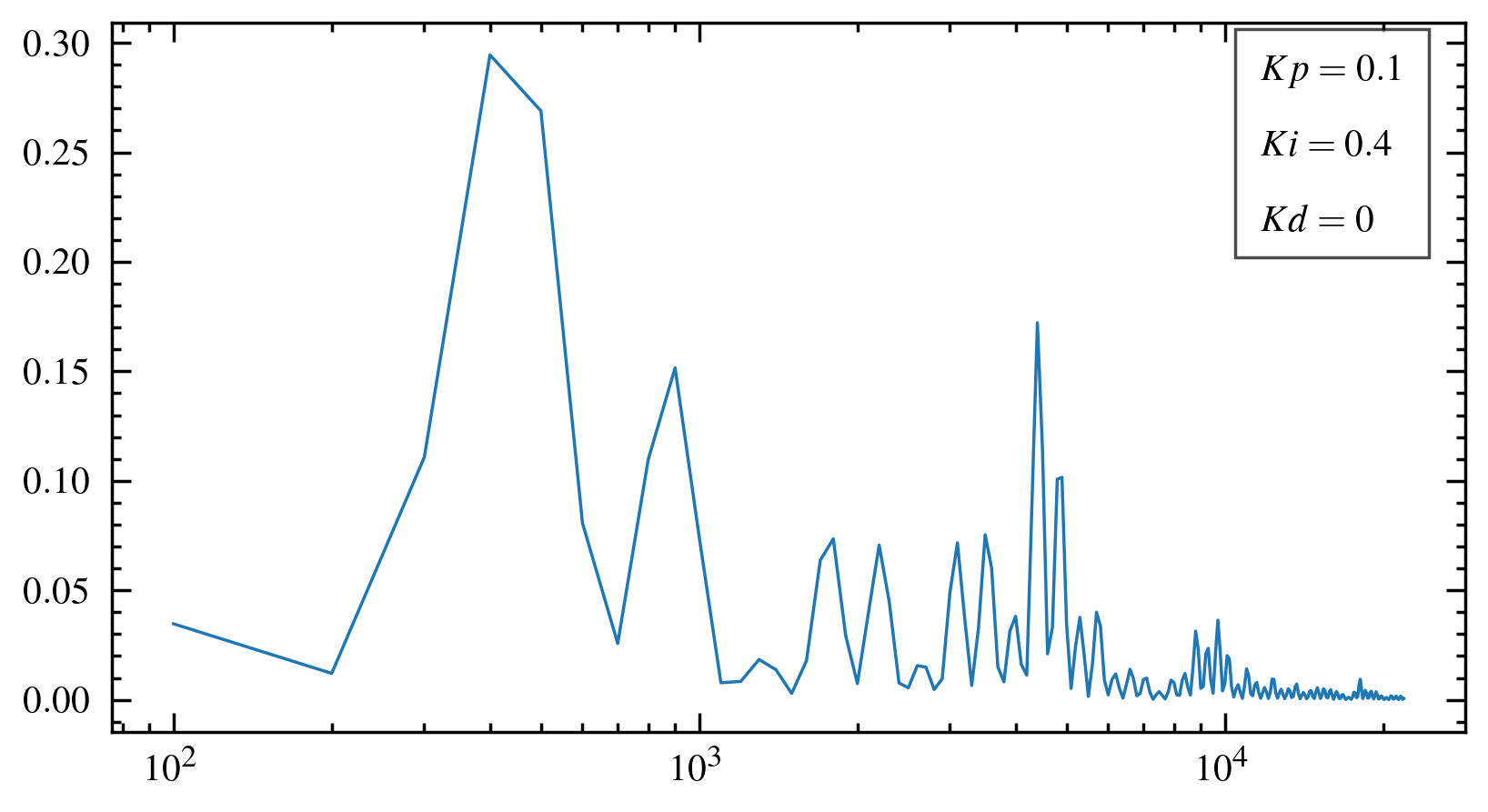}\label{F23b}}
\caption{The I-mode in PIDS can induce FM synthesis-like sidebands in the $output$ spectrum. In this case, varying $K_{i}$ with time is equivalent to the contribution of the FM modulation index.}
\label{F23}
\end{figure}

FM Synthesis is widely known to simulate naturally occurring timbres by adding side frequencies (harmonic and inharmonic) in the generated waveform and dynamically varying them with time \cite{chowning}. With the I-mode activated, PIDS can also exhibit similar behaviour. The I-component adds oscillations in the generated $output$, resulting in sidebands in the corresponding spectrum about the peak frequency; effectively, it distributes the energy of the peak among these sidebands. Additionally, the variation of $K_{i}$ (and other PIDS parameters) can alter the spectrum with time to recreate the effect of changing the modulation index in FM synthesizers; as $K_{i}$ increases, the distribution of energy among the upper sideband increases. An example of a PIDS $output$ spectrum similar to those produced by FM synthesizers is exhibited in \cref{F23}.

An important thing to note in this case is that the P- and D-components tend to oppose this FM synthesis-like behaviour induced by the I-component. As $K_{p}$ and $K_{d}$ increases with respect to $K_{i}$, the distribution of energy among the sidebands decreases and begins to increasingly accumulate at the peak/fundamental frequency.

\subsection{Alternative to Wavetable Synthesis}
\label{S64}

\begin{figure}[b]
\centering
\subfloat[]{\includegraphics[width=0.24\textwidth]{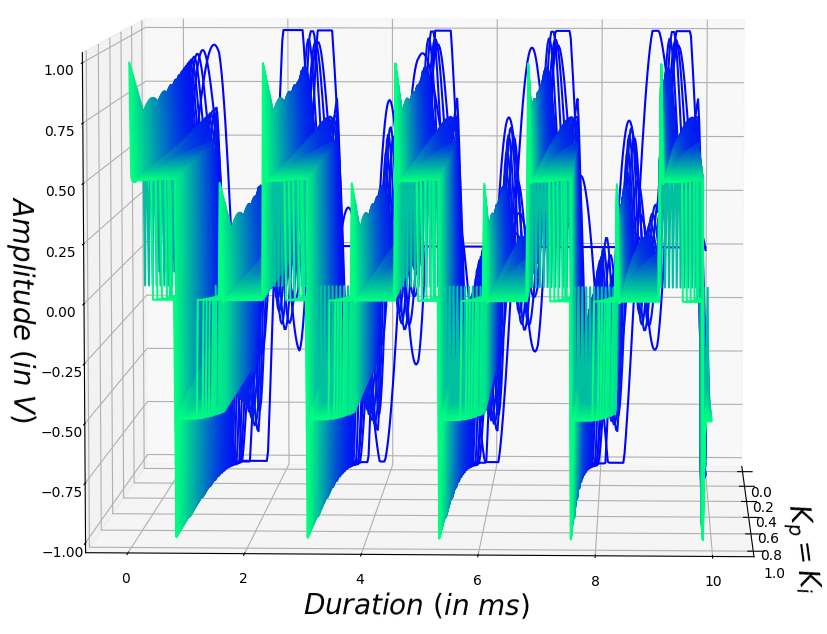}\label{F20a}}
\subfloat[]{\includegraphics[width=0.24\textwidth]{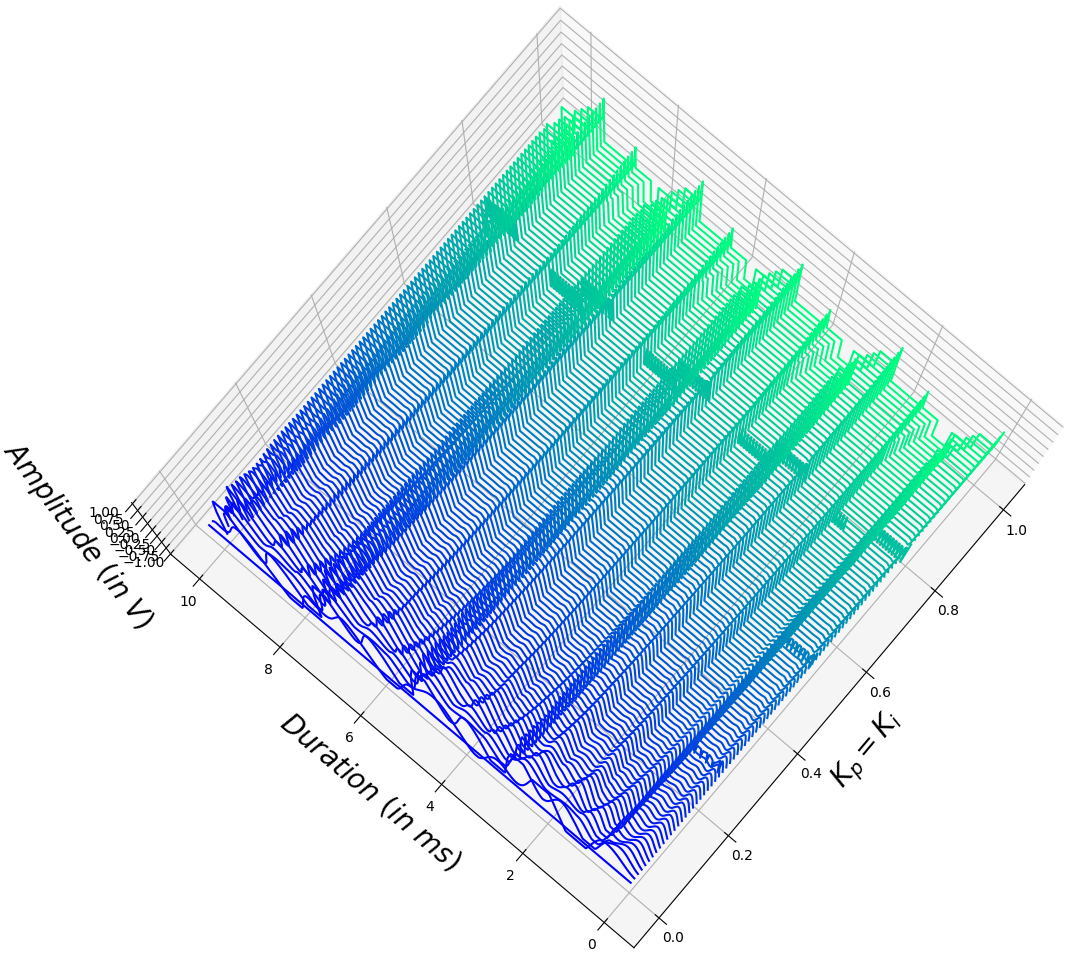}\label{F20b}}
\caption{By smoothly modulating one or more parameters over time, PIDS can generate $outputs$ similar to wavetable synthesis. Doing so doesn't require memory buffers and transitional interpolations between the "waves" in the table. Here, $K_{p}$ and $K_{i}$ are varied across time to obtain a wavetable transitioning from square waves (green side) to complex sinusoidal waveforms (blue side).}
\label{F20}
\end{figure}

Wavetable synthesis is one of the most popular forms of audio signal generation techniques in modern music. It involves modulating the selection of one of the available waveforms in a wavetable and interpolating between consecutive selections to smoothen the transitions \cite{bristow}. As a result, the process leads to intriguingly complex waveforms. The P, I and D components, as well as the breakpoint coordinates and type of the $artist$ used, greatly influence the resulting $output$ synthesized by PIDS. Hence, modulating one or more of these parameters can return an effect similar to modulating a wavetable.

To this end, PIDS presents some advantages over conventional wavetable synthesis. Firstly, all transitions in waveform shapes take place in situ. Hence, no wavetable buffer is required to be maintained in memory, thereby benefiting processors having low-memory or read speeds. Further, the number of waves in the PIDS "wavetable" is essentially a function of the sampling rate of the modulating signal. Due to the nature of PID control, the transitions between the waves will have inherent smoothness despite sudden changes in PIDS parameters, and thus, the need for digital interpolation is avoided.

There may be limitless possibilities from using PIDS as an alternative to wavetable synthesis. The influencing factors being the modulating signal used (could be conventional or PIDS LFOs) and the parameters being modulated. A simple instance of the "wavetable" generated by linearly sweeping the $K_{p}$ and $K_{i}$ parameters is seen in \cref{F20}.

\section{Further Research Opportunities}
\label{S7}

\subsection{Addressing unstable conditions}
\label{S71}

\begin{figure} [!b]
\centering
\subfloat[Time-domain plot of an unstable PIDS $output$.]{\includegraphics[width=0.24\textwidth]{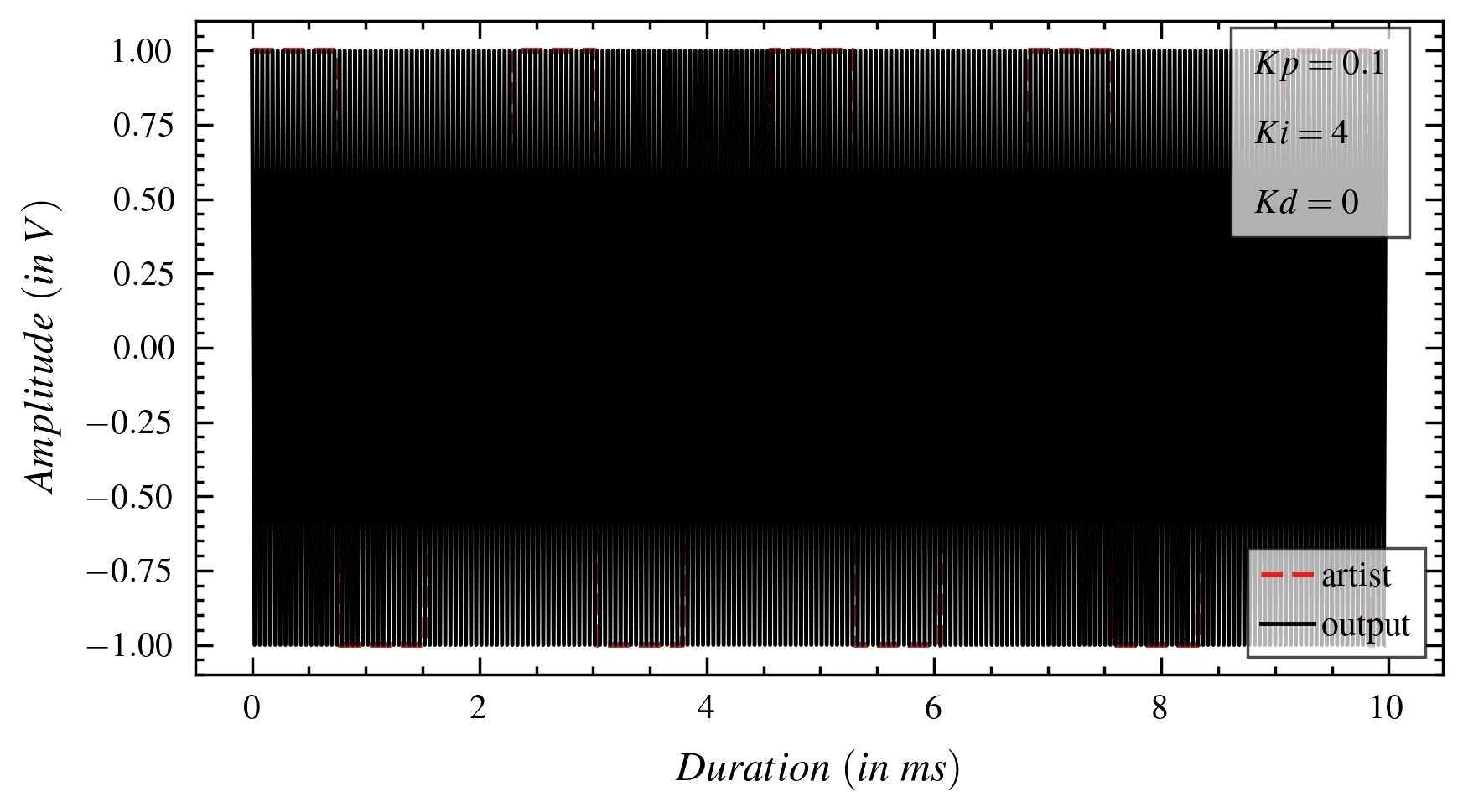}\label{F25a}}
\subfloat[Frequency-domain plot of an unstable PIDS $output$.]{\includegraphics[width=0.24\textwidth]{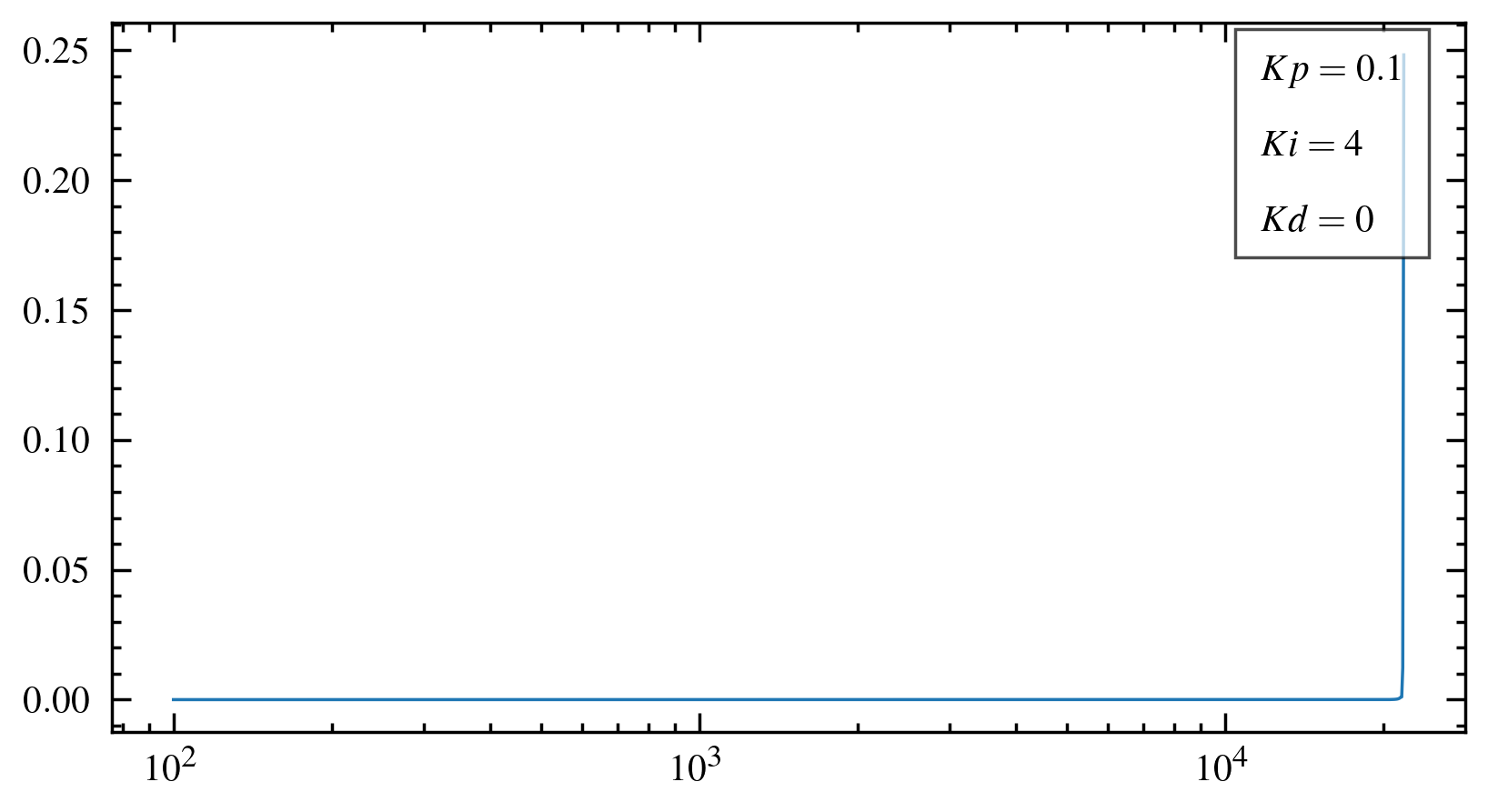}\label{F25b}}\hfill
\subfloat[Time-domain plot of an unstable PIDS $output$ after anti-aliasing.]{\includegraphics[width=0.24\textwidth]{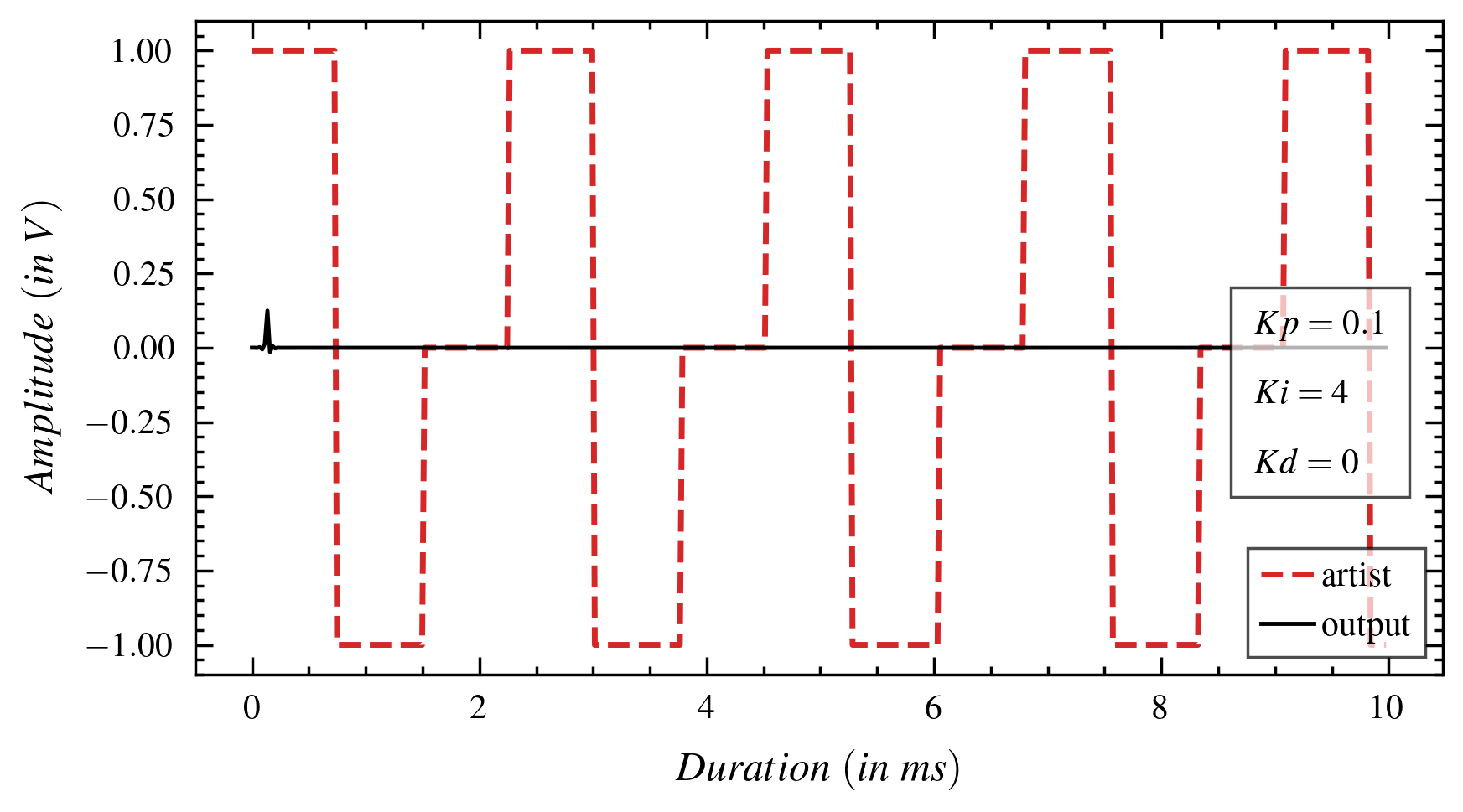}\label{F25c}}
\subfloat[Frequency-domain plot of an unstable PIDS $output$ after anti-aliasing.]{\includegraphics[width=0.24\textwidth]{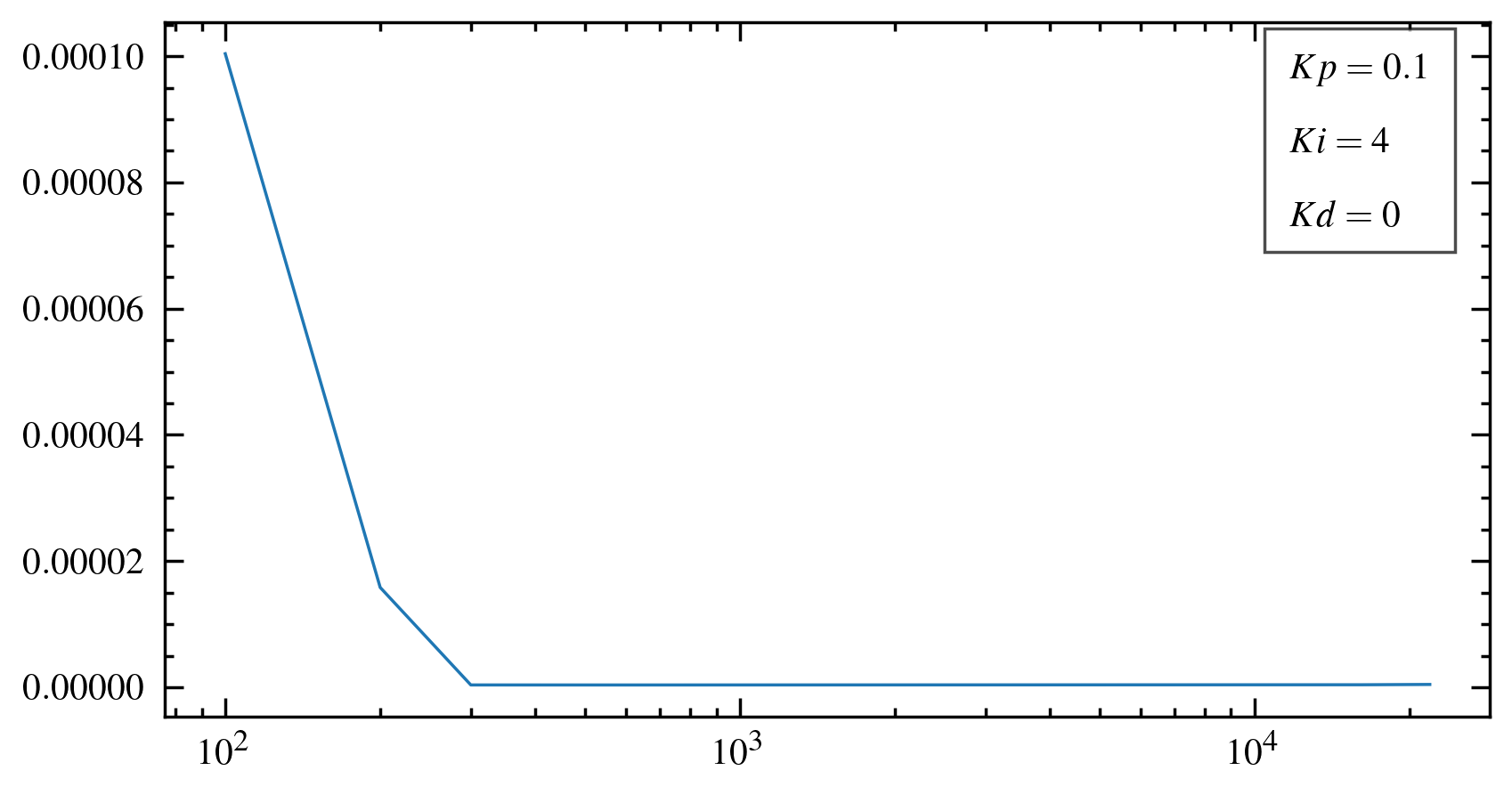}\label{F25d}}\hfill
\caption{As seen in \cref{F25a} and \cref{F25b} , setting the PID gain parameters to very high values tends to tip PIDS into instability, resulting in an uncontrollable triangle wave generated at Nyquist frequency. Anti-aliasing implementations will remove this high-frequency component in the signal resulting in an inaudible signal.}
\label{25}
\end{figure}

Similar to its applications in the process control industry \cite{liptak}, the PID control scheme in PIDS is also susceptible to operating as an unstable control system. Such situations occur when one or more $K_{p}$, $K_{i}$ and $K_{d}$ values are very high, leading the $output$ to oscillate continuously between two values (i.e., a triangular waveform) at the Nyquist frequency. Ultimately, any anti-aliasing mechanism incorporated into PIDS tends to filter out this component completely, resulting in an inaudible signal like in the case of \cref{F25b}. As of current progress in this research, it has not proven easy to estimate the exact gain values at which such conditions occur. A simple means to reduce the chances of instability may be to limit the range of inputs set for $K_{p}$, $K_{i}$ and $K_{d}$; the range of [0,1] is a convenient option. However, such range limiting may not be sufficient since, for instance, PIDS may operate stably at a particular value of $K_{p}$ less than one but may undergo instability for the same value when $K_{i}$ is activated. Thus, an onus is placed on the user to ensure that the parameter inputs they provide do not tip over PIDS into instability. Special care must be taken during the automation of these parameters to ensure that PIDS is stable at all instances of modulation. However, a long term solution must be found to objectively and completely remove the possibility of unstable conditions in PIDS.

\subsection{Addressing the DC Component}
\label{S72}

\begin{figure}[b]
\centering
\subfloat[Time-domain]{\includegraphics[width=0.24\textwidth]{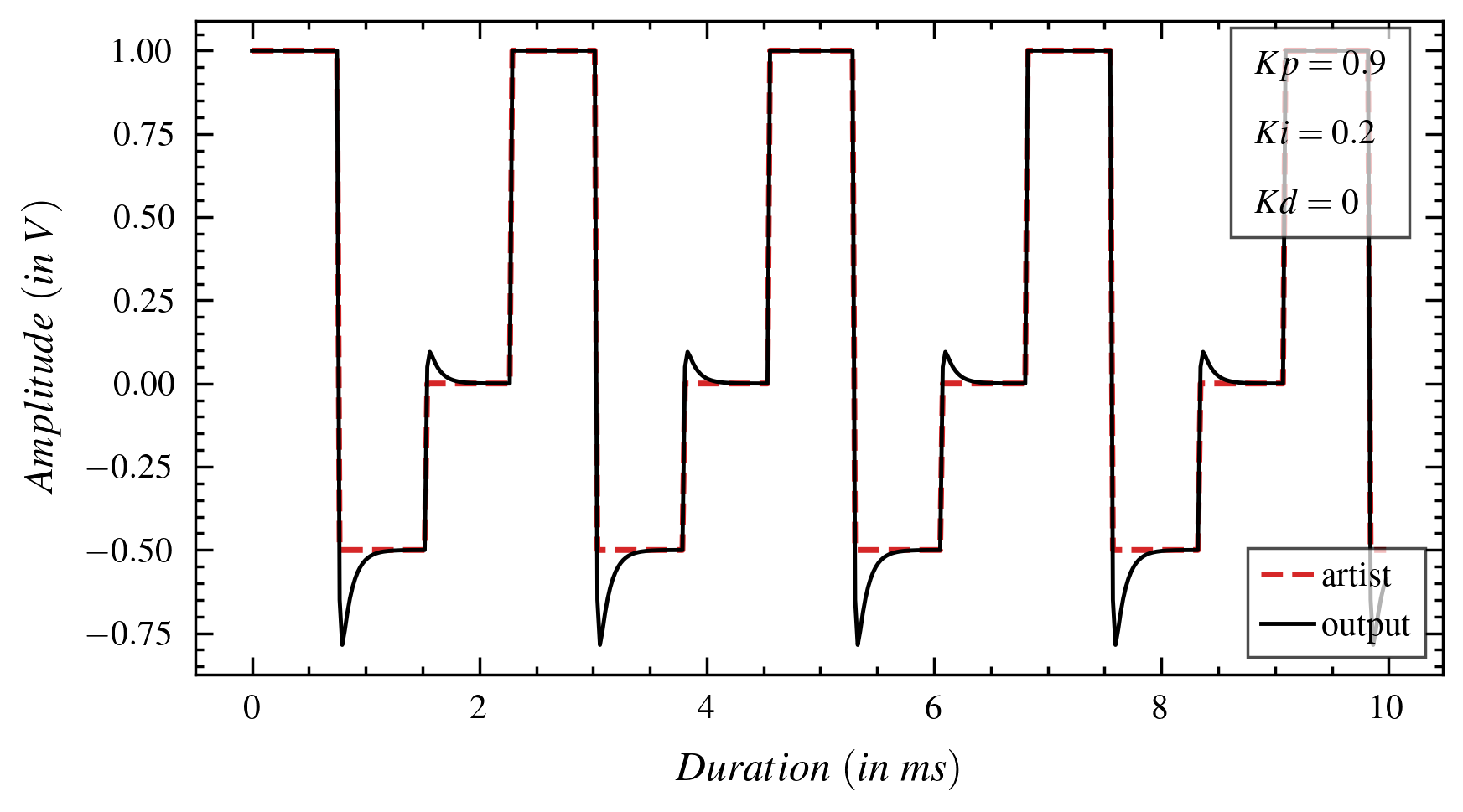}\label{F21a}}
\subfloat[Frequency-domain]{\includegraphics[width=0.24\textwidth]{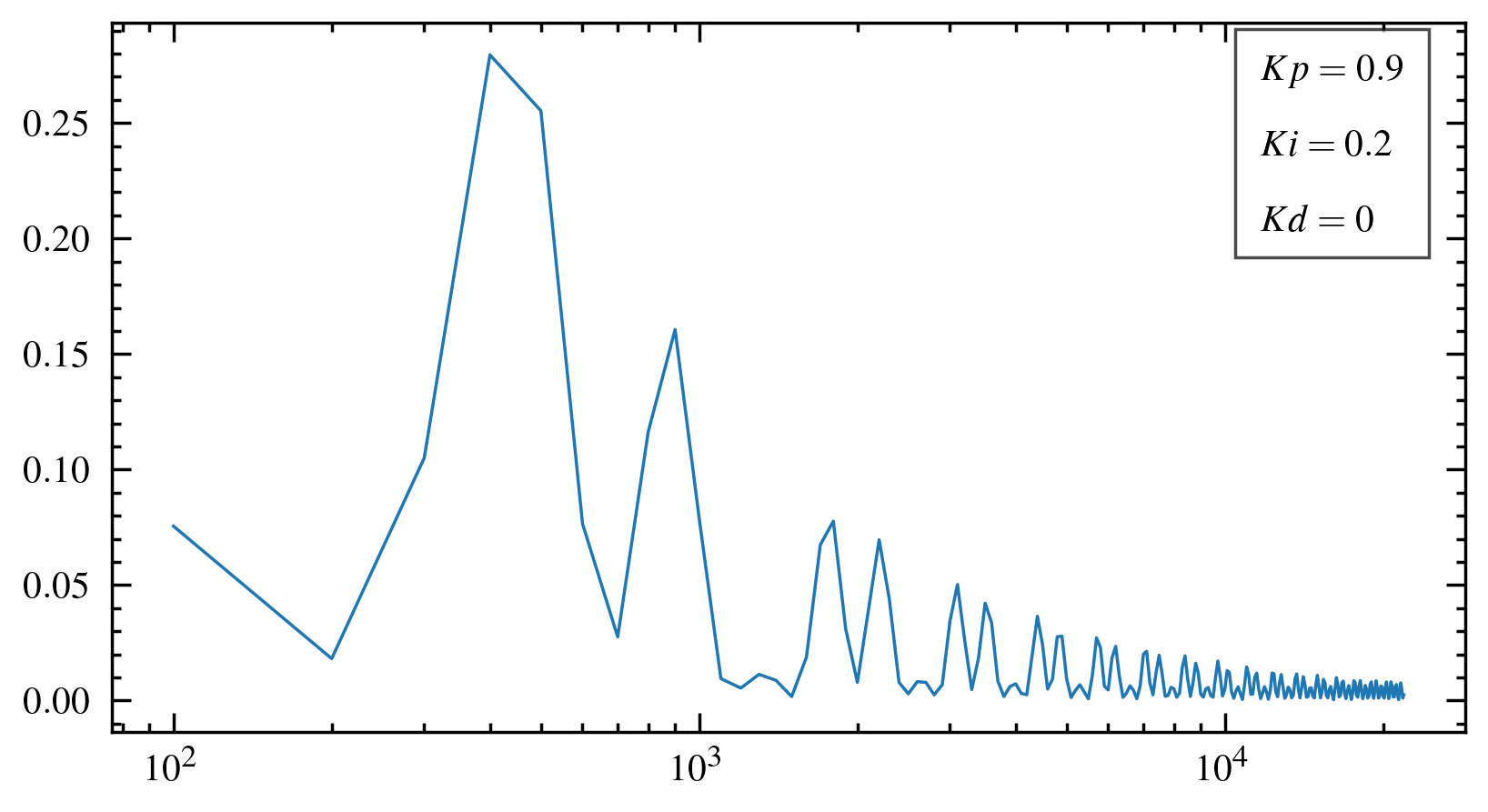}\label{F21b}}
\caption{When PIDS $output$ is symmetrical about 0V, a significant low-frequency component is obtained on its frequency spectrum known as the DC component.}
\label{F21}
\end{figure}

Unlike conventional synthesizers, the PIDS algorithm cannot guarantee that the synthesized audio waveforms will be symmetric about the time axis. A significant low-frequency component is present in the spectrum for signals having asymmetries, known as the DC component or DC bias of that signal, as seen in \cref{F21}. The DC component can be a source of unwanted "click" sounds and distortions that may be amplified when using some effect processors or while exporting the signal in the MP3 format \cite{dcoffset}. Hence, there may be an interest in removing it from the synthesized $output$. Some possible solutions in this regard may be implementing a high pass filter or subtracting the moving average of the $output$ from its instantaneous value. However, this study did not employ these strategies to address the DC component as it modifies the actual PIDS produced $output$, and there may be better solutions available to remove it intrinsically.

\subsection{Identifying Intrinsic Anti-Aliasing Techniques}
\label{S73}

In the aforementioned PIDS framework, anti-aliasing is achieved by oversampling. While this strategy is sufficient in practical cases to reduce aliasing to a great extent, it seems unnatural and detached from the PID control technique. Additionally, the computational complexity of this method scales up with an increase in the magnitude of oversampling (that may be required for the higher audio frequencies) and the order of the filter used \cite{kahles}.

\begin{figure}
\centering
\includegraphics[width=0.48\textwidth]{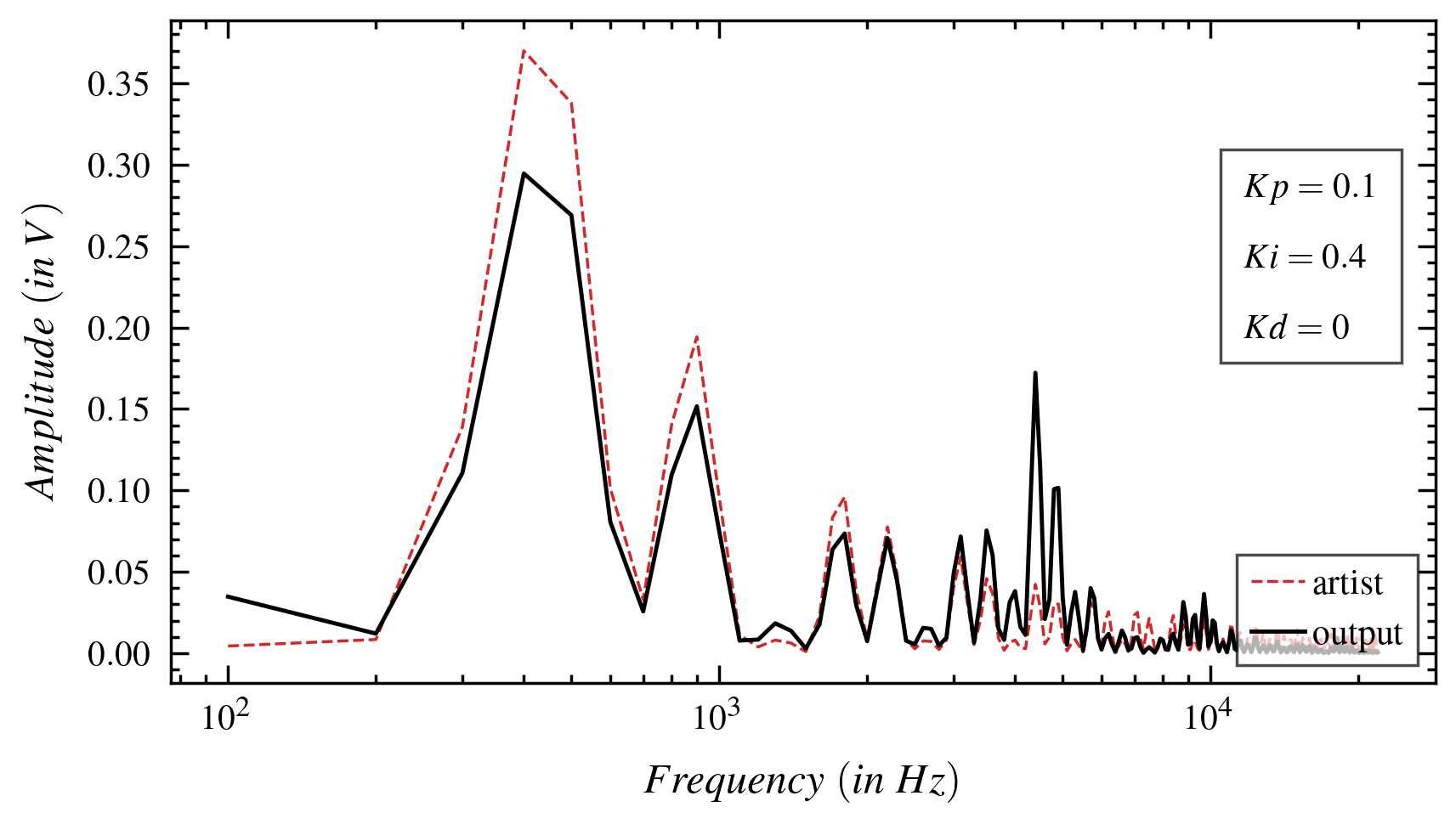}
\caption{Most of the typical $artist$ types contain high-frequency components in their spectrum. While PIDS can remove a considerable part of these components from the synthesized $output$, the process itself adds distinct high-frequency features of its own. Since these components can lead to aliasing, a convenient and intrinsic technique of anti-aliasing must be incorporated into the PIDS algorithm to bypass the need for oversampling.}
\label{F22}
\end{figure}

In audio waveforms, a significant source of high-frequency components prone to aliasing are discontinuities in the form of sharp changes in magnitudes (found in square and sawtooth waves) or slope (found in triangle waves). This is true for the linear and step $artists$ against which PIDS $output$ is synthesized. The PID algorithm can behave as a lag-compensator and negate the effect of high-frequency components in the $artist$. However, the algorithm also adds its own high frequencies in the $output$. \Cref{F22} represents these observations.

Therefore, effort must be made to identify strategies to reduce the PID induced aliased components in the $output$ and capitalize on its intrinsic low pass filtering tendencies to execute a form of anti-aliasing naturally integrated into the PID control mechanism. As a starting point in this regard, implementing a small fraction of the integral mode (even when $K_{i}$ is zero) may be considered.

\section{Conclusion}
\label{S8}

This study discussed the framework for a novel audio synthesis technique derived from the fundamentals of the widely used PID controllers in the process industry. Here, the goal was to lay out the foundation of the synthesis technique in terms of the nature of synthesized $outputs$ that may be produced. Additionally, the concept of $artists$ and their constituent breakpoints was introduced as a significant building block of the PIDS framework.

Efforts were also made to understand the effects of each control parameter on the synthesized $output$; these included $K_{p}$, $K_{i}$ and $K_{d}$, as well as the type of $artist$ and the number of breakpoints used. From a practical perspective, the problems of aliasing and some common ways to handle them were analyzed.

Finally, the research delved into a high-level demonstration of the possible applications of PIDS. It is conducive to being used in additive synthesis and also as an LFO generator. It also presents itself as a capable alternative to FM and wavetable synthesis, emulating their effects with efficiency.

However, there are aspects of PIDS that call for concentrated research efforts. Identifying techniques to prevent instability and developing intrinsic forms of anti-aliasing are some of the future research directions required before employing PIDS in commercial synthesizers. Moreover, additional experimentation may be needed from a musician's perspective to fully comprehend the possible applications of PIDS in real-world music production and performance.

As of now, PIDS is just a newfound, promising technique in audio synthesis. Nevertheless, with a focused approach to address the imperfections and optimize the implementation, it can influence future developments in generating musical and non-musical sounds and control signals.

\bibliographystyle{IEEEtran}
\bibliography{pido_paper_citations.bib}

\end{document}